\documentclass[11pt,letter]{article}

\usepackage{microtype}
\usepackage{graphicx}
\usepackage{epstopdf}
\usepackage{xspace,enumerate,enumitem,boxedminipage,fullpage}
\usepackage{algorithm,algorithmic}
\usepackage{mathtools,amsmath,amssymb,amsfonts}
\usepackage{color,colortbl}

\usepackage{rotating}
\usepackage[mathscr]{euscript}		

\setlist[description]{listparindent=\parindent,leftmargin=0em,font=\textbf,itemsep=0.5em}

\usepackage{perpage}
\MakePerPage{footnote}

\newcommand{\ie}{\emph{i.e.}}
\newcommand{\eg}{\emph{e.g.}}
\newcommand{\etal}{\emph{et al.}}
\newcommand{\alg}[1]{\mbox{\textsf{#1}}}
\newcommand{\func}[1]{\mbox{$F_\textsf{#1}$}}
\newcommand{\simu}[1]{\mbox{$\mathcal{S}_\textsf{#1}$}}

\newcommand{\mask}[1]{\widehat{#1}}

\newcommand{\e}{\mathrm{e}}
\newcommand{\F}{\mathbb{F}}

\DeclareMathOperator{\polylog}{polylog}
\DeclareMathOperator{\poly}{poly}
\DeclareMathOperator*{\Prob}{Prob}
\floatname{algorithm}{Protocol}

\usepackage{amsthm}
\newtheoremstyle{mytheoremstyle}
	{\topsep}                    
	{\topsep}                    
	{\itshape}                   
	{}                           
	{\bfseries}                   
	{.}                          
	{0.5em}                       
	{}  

\theoremstyle{mytheoremstyle}
\newtheorem{theorem}{Theorem}
\newtheorem{lemma}{Lemma}
\newtheorem{corollary}{Corollary}

\floatname{algorithm}{Protocol}

\let\savedCaption=\caption
\renewcommand*{\caption}[1]{\savedCaption[#1]{~#1}}  

\usepackage{caption}

\newcommand{\algfont}{}

\newcommand{\fbad}{1/8\xspace}
\newcommand{\fbadq}{$q/8$\xspace}
\newcommand{\fgoodq}{7/8\xspace}
\newcommand{\fgoodqq}{$\frac{7q}{8}$\xspace}

\newcommand{\threshq}{$\frac{5q}{8}$\xspace}
\newcommand{\dcl}{D}
\newcommand{\bins}{k}
\newcommand{\flag}{$\langle\mathsf{Flag}\rangle$\xspace}

\newcommand{\countm}{$\langle\mathsf{Count}\rangle$\xspace}
\newcommand{\done}{$\langle\mathsf{Done}\rangle$\xspace}
\newcommand{\oneinput}{$\mathsf{1}$-input\xspace}
\newcommand{\oneinputs}{$\mathsf{1}$-inputs\xspace}

\newcommand{\vssSh}{\textsf{VSS-Share}\xspace}
\newcommand{\vssRec}{\textsf{VSS-Reconst}\xspace}
\newcommand{\hw}{\mbox{\textsf{CMPC}}\xspace}
\newcommand{\ba}{\mbox{\textsf{BA}}\xspace}

\newcommand{\ce}{\textsf{Circuit-Eval}\xspace}

\newcommand{\ct}{count tree\xspace}

\newcommand{\toplev}{\ell^*}

\newcommand{\lev}{\ell}

\newcommand{\Sendhop}{\textsf{Sendhop}}
\newcommand{\Send}{\textsf{Send}}
\newcommand{\MessagePass}{\textsf{MessagePass}}

\newcommand{\heavyba}{{byzantine agreement}\xspace}
\newcommand{\lst}{\ensuremath{\mathsf{List}}}
\newenvironment{indentpar}[1]%
{\begin{list}{}%
	{\setlength{\leftmargin}{#1}}%
	\item[]%
}
{\end{list}}

\newcommand{\rs}{semi-random-string agreement\xspace}
\newcommand{\binelection}{{bin election}\xspace}
\newcommand{\es}{\mbox{\textsf{Elect-Subcommittee}}\xspace}
\newcommand{\simplees}{\mbox{\textsf{Simple-Elect-Subcommittee}}\xspace}
\newcommand{\rsAlg}{\mbox{\textsf{SRS-Agreement}}\xspace}
\newcommand{\rsToQ}{\mbox{\textsf{SRS-to-Quorum}}\xspace}

\newcommand{\papertitle}{Secure Multi-Party Computation in Large Networks}
\definecolor{Gray}{gray}{0.92}
\definecolor{darkblue}{rgb}{0,0,0.4}

\newcounter{claimcounter}
\newenvironment{claim}{\stepcounter{claimcounter}{\vspace{0.5em}\noindent \textit{Claim \theclaimcounter.~}}}{}

\usepackage{authblk}

\author[1]{\normalsize Varsha Dani\thanks{varsha@cs.unm.edu}}
\author[2]{Valerie King\thanks{val@uvic.ca}}
\author[1]{Mahnush Movahedi\thanks{movahedi@cs.unm.edu}}
\author[1]{Jared Saia\thanks{saia@cs.unm.edu}}
\author[1]{Mahdi Zamani\thanks{zamani@cs.unm.edu}}
\affil[1]{Department of Computer Science, University of New Mexico, Albuquerque, NM, USA\vspace{3pt}}
\affil[2]{Department of Computer Science, University of Victoria, Victoria, BC, Canada}

\begin{document}
\title{\papertitle\thanks{This is the extended version of the paper published in the proceedings of the 2014 International Conference on Distributed Computing and Networking (ICDCN 2014). This research was partially supported by NSF CAREER Award 0644058 and NSF grants CCR-0313160 and CCF-1320994.}}
	
\sloppy		

\maketitle
\begin{abstract}
We describe scalable protocols for solving the secure multi-party computation (MPC) problem among a large number of parties. We consider both the synchronous and the asynchronous communication models. In the synchronous setting, our protocol is secure against a static malicious adversary corrupting less than a $1/3$ fraction of the parties. In the asynchronous setting, we allow the adversary to corrupt less than a $1/8$ fraction of parties. For any deterministic function that can be computed by an arithmetic circuit with $m$ gates, both of our protocols require each party to send a number of field elements and perform an amount of computation that is $\tilde{O}(m/n + \sqrt n)$. We also show that our protocols provide perfect and universally-composable security.

\vspace{-1pt}
To achieve our asynchronous MPC result, we define the \emph{threshold counting problem} and present a distributed protocol to solve it in the asynchronous setting. This protocol is load balanced, with computation, communication and latency complexity of $O(\log{n})$, and can also be used for designing other load-balanced applications in the asynchronous communication model.
\end{abstract}

\section{Introduction}
In \emph{secure multi-party computation (MPC)}, a set of parties, each having a secret value, want to compute a common function over their inputs, without revealing any information about their inputs other than what is revealed by the output of the function.
Recent years have seen a renaissance in MPC, but unfortunately, the distributed computing community is in danger of missing out. In particular, while new MPC algorithms boast dramatic improvements in latency and communication costs, none of these algorithms offer significant improvements in the highly \emph{distributed} case, where the number of parties is large.

This is unfortunate, since MPC holds the promise of addressing many important problems in distributed computing. How can peers in BitTorrent auction off resources without hiring an auctioneer? How can we design a decentralized Twitter that enables provably anonymous broadcast of messages. How can we create deep learning algorithms over data spread among large clusters of machines?

Most large-scale distributed systems are composed of nodes with limited resources. This makes it of extreme importance to \emph{balance} the protocol load across all parties involved. Also, large networks tend to have weak admission control mechanisms which makes them likely to contain malicious nodes. Thus, a key variant of the MPC problem that we consider will be when a certain hidden fraction of the nodes are controlled by a malicious adversary.

\subsection{Our Contribution} \label{sec:contribution}
In this paper, we describe general MPC protocols for computing arithmetic circuits. In terms of communication and computation costs per party, our protocols scale sublinearly with the number of parties and linearly with the size of the circuit. 

To achieve sublinear communication and computation costs, our protocols critically rely on the notion of \emph{quorums}. A quorum is a set of $O(\log{n})$ parties, where the number of corrupted parties in each quorum is guaranteed not to exceed a certain fraction. We describe an efficient protocol for creating a sufficient number of quorums in the asynchronous setting.

To adapt to the asynchronous setting, we introduce the general problem of \emph{threshold counting}. We show how this problem relates to the problem of dealing with arbitrarily-delayed inputs in our asynchronous MPC protocol, and then propose an efficient protocol for solving it.

When a protocol is concurrently executed alongside other protocols (or with other instances of the same protocol), one must ensure this composition preserves the security of the protocol. We show that our protocols are secure under such concurrent compositions by proving its security in the \emph{universal composability (UC) framework} of Canetti~\cite{Canetti:UCSecurity:2001}.

\subsection{Model} \label{sec:model}
Consider $n$ parties $P_1,...,P_n$ in a fully-connected network with private and authenticated channels.
In our asynchronous protocol, we assume communication is via asynchronous message passing, so that sent messages may be arbitrarily and adversarially delayed. Latency (or running time) of a protocol in this model is defined as the maximum length of any chain of messages sent/received throughout the protocol (see \cite{CD,Attiya:DC:2004}).

We assume a \emph{malicious} adversary who controls an unknown subset of parties. We refer to these parties as \emph{corrupted} and to the remaining as \emph{honest}. The honest parties always follow our protocol, but the corrupted parties not only may share information with other corrupted parties but also can deviate from the protocol in any arbitrary manner, \eg, by sending invalid messages or remaining silent. 

We assume the adversary is \emph{static} meaning that it must select the set of corrupted parties at the start of the protocol. We assume that the adversary is computationally-unbounded; thus, we make no cryptographic hardness assumptions.

\subsection{Problem Statement} \label{sec:problem}
\begin{description}
	\item[Multi-Party Computation.]
	In the MPC problem, $n$ parties, each holding a private input, want to jointly evaluate a deterministic $n$-ary function $f$ over their inputs while ensuring:
	
	\smallskip
	\begin{enumerate}
		\item Each party learns the correct output of $f$; and
		\item No party learns any information about other parties' inputs other than what is revealed from the output.
	\end{enumerate}
	
	\item[Constraints for the Asynchronous Model.]
	Consider a simple setting, where, the $n$ parties send their inputs to a trusted party $P$ who then locally computes $f$ and sends the result back to every party.
	In the asynchronous setting, the MPC problem is challenging even with such a trusted party. In particular, since the $t$ corrupted parties can refrain from sending their inputs to $P$, it can only wait for $n-t$ inputs rather than $n$ inputs. Then, it can compute $f$ over $n$ inputs consisting of $n-t$ values received from the parties and $t$ dummy (default) values for the missing inputs. Finally, the trusted party sends the output back to the parties.
	The goal of asynchronous MPC is to achieve the same functionality as the above scenario but without the trusted party $P$.
	
	discuss the termination condition based FLP impossibility. In this paper, we show that protocol terminates with high probability.  
	
	\item[Quorum Building.]
	A quorum is a set of $\Theta(\log{n})$ parties, where the fraction of corrupted parties in this set is at most $t/n+\epsilon$ for a small positive constant $\epsilon$. In the quorum building problem, there are $n$ parties up to $t < n$ of whom may be corrupted. The goal is to ensure all parties agree on a set of $n$ quorums such that each party is mapped to $O(\log{n})$ quorums.

	\item[Threshold Counting.]
	In this problem, there are $n$ honest parties each with a flag bit initially set to $0$. At least $\tau < n$ of the parties will eventually set their bits to $1$. The goal is for all the parties to learn when the number of bits set to $1$ becomes greater than or equal to $\tau$.
	
\end{description}


\subsection{Our Results} \label{sec:results}
The main results of this paper are summarized by the following theorems proved in Section~\ref{sec:mainproofs}. We consider an $n$-ary function, $f$, represented as an arithmetic circuit of depth $d$ with $m$ gates. 

\begin{theorem} \label{thm:main-sync}
	There exist a universally-composable protocol that with high probability solves the synchronous MPC problem and has the following properties:
	\begin{itemize}
		\item It is secure against $t < (1/3-\epsilon)n$ corrupted parties, for some fixed $\epsilon>0$.
		\item Each party sends $\tilde{O}(m/n + \sqrt n)$ field elements.
		\item Each party performs $\tilde{O}(m/n + \sqrt n)$ computations.
		\item The expected running time is $O(d \polylog(n))$.
	\end{itemize}
\end{theorem}


\begin{theorem} \label{thm:main}
	There exist a universally-composable protocol that with high probability solves the asynchronous MPC problem and has the following properties:
	\begin{itemize}
		\item It is secure against $t < (1/8-\epsilon)n$ corrupted parties, for some fixed $\epsilon>0$.
		\item Each party sends $\tilde{O}(m/n + \sqrt n)$ field elements.
		\item Each party performs $\tilde{O}(m/n + \sqrt n)$ computations.
		\item The expected running time is $O(d \polylog(n))$.
	\end{itemize}
\end{theorem}

\begin{description}
\item[Paper Organization.]
In Section~\ref{sec:relatedwork}, we discuss related work. In Section~\ref{sec:prelim}, we define our notation and discuss the building blocks used in our protocols. We present our MPC protocols in Section~\ref{sec:alg}. In Section~\ref{sec:mainproofs}, we prove the security of our MPC protocols. Section~\ref{sec:taucount} is a self-contained presentation of the threshold counting problem and our solution to this problem. In Section~\ref{sec:qf}, we describe an asynchronous protocol for the quorum building problem. Finally, we conclude in Section~\ref{sec:end} and discuss future directions.
\end{description}

\section{Related Work} \label{sec:relatedwork}
Due to the large body of work, we do not attempt a comprehensive review of the MPC literature here, but rather focus on seminal work and, in particular, schemes that achieve sublinear per-party communication costs.
The MPC problem was first described by Yao~\cite{Yao:1982:PSC:1382436.1382751}. He described an algorithm for MPC with two parties in the presence of a semi-honest adversary. Goldreich~\etal~\cite{Goldreich:1987:PAM:28395.28420} propose the first MPC protocol that is secure against a malicious adversary. This work along with~\cite{Chaum:1987:MCE:646752.704756,Galil:1987:CCS:646752.704741} are all based on cryptographic hardness assumptions. 
These were later followed by several cryptographic improvements~\cite{Beaver:1990:RCS:100216.100287,Gennaro:1998:SVF:277697.277716,Canetti:1996:ASM:888604}.

In a seminal work, Ben-Or~\etal~\cite{bgw88} show that every function can be computed with information-theoretic security in the presence of a semi-honest adversary controlling less than half of the parties, and in the presence of a malicious adversary controlling less than a third of the parties. They describe a protocol for securely evaluating an arithmetic circuit that represents the function. 

This work was later improved in terms of both communication and computation costs in \cite{chaum_crepeau_damgard:multiparty,Beaver:1991,Gennaro:1998:SVF:277697.277716}. Unfortunately, these methods all have poor communication scalability. In particular, if there are $n$ parties involved in the computation, and the function $f$ is represented by a circuit with $m$ gates, then these algorithms require each party to send a number of messages and perform a number of computations that is $\Omega(nm)$.

These were followed by several improvements to the cost of MPC, when $m$ (\ie, the circuit size) is much larger than $n$~\cite{damgard2006scalable,damgard2007scalable,damgard2008scalable}. For example, the protocol of Damg{\aa}rd~\etal~\cite{damgard2008scalable} incurs computation and communication costs that are $\tilde{O}(m)$ plus a polynomial in $n$. Unfortunately, the additive polynomial in these algorithms is large (at least $\Omega(n^{6})$) making them impractical for large $n$. One may argue that for large circuits the circuit-dependent complexity dominates the polynomial complexity. However, we believe there are many useful circuits such as the ones used in~\cite{msz:sirocco:2015,Hamada:2012:MPCSorting} which have relatively small number of gates.

\begin{description}
	\item[Asynchronous MPC.]
	Foundational work in asynchronous MPC was presented by Ben-Or~\etal~\cite{benor_canetti_goldreich:asynchronous}. They adapt the protocol of \cite{bgw88} to the asynchronous setting and show that asynchronous MPC is possible for up to $n/3$ fail-stop faults and up to $n/4$ malicious faults. Improvements were made by Srinathan and Rangan~\cite{srinathan_pandu_rangan:efficient} and Prabhu~\etal~\cite{prabhu_srinathan_rangan:asynchronous} with a final
	communication cost of $O(n^3)$ per multiplication achieved by Beerliov\'{a}-Trub\'{\i}niov\'{a} and Hirt~\cite{Beerliova-Trubiniova:2007:SEP:1781454.1781486} for perfectly-secure asynchronous MPC with the optimal resiliency bound of up to $n/4$.
	
	Damg{\aa}rd~\etal~\cite{Damgard:2009:AMC:1531954.1531967} describe a perfectly-secure MPC that guarantees termination only when the adversary allows a preprocessing phase to terminate. However, their protocol is not fully asynchronous, as they assume a few synchronization points; hence, they can achieve a resiliency bound of up to $n/3$.
	
	Choudhury~\etal~\cite{Choudhury:2013:AsyncMPC} propose an amortized asynchronous MPC protocols with linear communication complexity per multiplication gate meaning that the communication done by an individual party for each gate does not grow with the number of parties. This protocol is unconditionally-secure against up to $n/4$ corrupted parties with a small failure probability. In our paper, we are directly addressing the third open problem of~\cite{Choudhury:2013:AsyncMPC} as we quote here: 
	
	\emph{``If one is willing to reduce the resilience t from the optimal resilience by a constant fraction, then by using additional techniques like packed secret sharing, committee election and quorum forming, one can achieve additional efficiency in the synchronous MPC protocols, as shown in [...]. It would be interesting to see whether such techniques can be used in the asynchronous settings to gain additional improvements.''}
	
	
	
	
	\item[MPC with Sublinear Overhead.]
	We first introduced the notion of using quorums to decrease message cost in MPC in a brief announcement~\cite{Dani:2012:BAB:2332432.2332473}. In that paper, we described a synchronous protocol with bit complexity of $\tilde{O}(m/n + \sqrt{n})$ per party that can tolerate a computationally unbounded adversary who controls up to $(1/3 - \epsilon)$ fraction of the parties for any fixed positive $\epsilon$. As network size scales, it becomes infeasible to require each party to communicate with all other parties. 
	
	The current paper is the detailed version of our later extended abstract~\cite{DKMS-ICDCN-2014}, where we described algorithms to improve~\cite{Dani:2012:BAB:2332432.2332473} by handling asynchronous communication. One important challenge in the asynchronous communication model is to ensure that at least $n-t$ inputs are committed to, before the circuit evaluation. To address this issue we introduce and solve the \emph{threshold counting problem.}
	
	Boyle~\etal~\cite{Boyle:2013:CLS:2450206.2450227} describe a synchronous MPC protocol for evaluating arithmetic circuits. The protocol is computationally-secure against an adversary corrupting up to ($1/3-\epsilon$) fraction of parties, for some fixed positive $\epsilon$. Similar to~\cite{Dani:2012:BAB:2332432.2332473}, the protocol of~\cite{Boyle:2013:CLS:2450206.2450227} also uses quorums to achieve sublinear per-party communication cost. Interestingly, the communication cost of this protocol is independent of circuit size. This is achieved by evaluating the circuit over encrypted values using a \emph{fully-homomorphic encryption (FHE)} scheme~\cite{Gentry:2009:FHE:1536414.1536440}. Unfortunately, the protocol is not fully load-balanced as it evaluates the circuit using only one quorum (called the supreme committee). The protocol requires each party to send $\mathsf{polylog}(n)$ messages of size $\tilde{O}(n)$ bits and requires $\mathsf{polylog}(n)$ rounds.
	
	Chandran~\etal~\cite{cryptoeprint:2014:615} address two limitations of the protocol of~\cite{Boyle:2013:CLS:2450206.2450227}: tolerating an adaptive adversary and achieving optimal resiliency (\ie, $t<n/2$ malicious parties). They replace the common reference string assumption of~\cite{Boyle:2013:CLS:2450206.2450227} with a different setup assumption called symmetric-key infrastructure, where every pair of parties share a uniformly-random key that is unknown to other parties. The authors also show how to remove the SKI assumption at a cost of increasing the communication locality by $O(\sqrt{n})$. Although this protocol provides small communication locality, the bandwidth cost seems to be super-polynomial due to large message sizes.
	
	
	Boyle~\etal~\cite{cryptoeprint:2014:404} describe a scalable technique for secure computation of RAM programs~\cite{Goldreich:1996:SPS:233551.233553} in large networks by performing local communications in quorums of parties. For securely evaluating a RAM program $\Pi$, their protocol incurs a total communication and computation of $\mathsf{poly}(n) + \tilde{O}(Time(\Pi))$ while requiring $\tilde{O}(|x| + Space(\Pi)/n)$ memory per party, where $Time(\Pi)$ and $Space(\Pi)$ are time and space complexity of $\Pi$ respectively, and $|x|$ denotes the input size.
	
	In Table~\ref{tab:SublinearMPC}, we review recent MPC results that provide sublinear communication locality. All of these results rely on some quorum building technique for creating a set of quorums each with honest majority.

\newcolumntype{L}[1]{>{\columncolor{Gray}\raggedright\let\newline\\\arraybackslash}p{#1}}
\newcolumntype{C}[1]{>{\centering\let\newline\\\arraybackslash}m{#1}}
\newcolumntype{R}[1]{>{\raggedleft\let\newline\\\arraybackslash}m{#1}}
\newcommand{\beforegap}{\rule{0px}{17px}}
\newcommand{\aftergap}{\\[7px] \hline}
\begin{table*}
	\centering
	\scriptsize
	\renewcommand{\arraystretch}{1.5}
	\caption{Recent MPC results with sublinear communication costs}
	\label{tab:SublinearMPC}
	\vspace{1em}
	\begin{tabular}
		{|L{4.3em}|C{3.3em}|C{4.5em}|C{2.7em}|C{4.5em}|C{7em}|C{7em}|C{5.5em}|C{5.5em}|C{3.9em}|}
		
		\rowcolor{Gray}
		
		\hline
		\bfseries Protocol &
		\bfseries Security &
		\bfseries Resiliency Bound &
		\bfseries Async? &
		\bfseries Assumes Broadcast Channel? &
		\bfseries Total Message Complexity &
		\bfseries Total Computation Complexity &
		\bfseries Latency &
		\bfseries Msg Size &
		\bfseries Load- Balanced?
		\aftergap
		
		\beforegap
		\cite{Boyle:2013:CLS:2450206.2450227} &	
		Crypto & 				
		$(1/3-\epsilon)n$ & 	
		No & 					
		No & 					
		$\tilde{O}(n)$ & 		
		$\tilde{\Omega}(n) + \tilde{\Omega}(\kappa m d^3)^\dag$ & 					
		$\tilde{O}(1)$ & 		
		$O(n \ell \cdot \mathsf{polylog}(n))$ & 		   
		No                 
		\aftergap
				
		\beforegap
		\cite{cryptoeprint:2014:404} &			 		
		Perfect & 				
		$(1/3-\epsilon)n$ & 	
		No & 					
		Yes & 					
		$\mathsf{poly}(n) + \tilde{O}\big(Time(\Pi)\big)$ & 	
		$\mathsf{poly}(n) + \tilde{O}\big(Time(\Pi)\big)$ & 	
		$\tilde{O}\big(Time(\Pi)\big)$ & 					
		$O(\ell)$ &					
		Yes                 
		\aftergap			
		
		\beforegap
		\cite{cryptoeprint:2014:615} &			 		
		Crypto$^\ddag$ & 					
		$n/2$ & 					
		No & 					
		No & 					
		$O(n\log^{1+\epsilon}{n})$ \newline or \newline $O(n\sqrt{n}\log^{1+\epsilon}{n})$ & 					
		$\Omega(n\log^{1+\epsilon}{n})$ \newline or \newline $\Omega(n\sqrt{n}\log^{1+\epsilon}{n})$ & 					
		$O(\log^{\epsilon^{\prime}}{n})$ & 					
		$\Omega\big(\log^{\log{n}}{n}\big)$ \newline or \newline $\Omega\big(\sqrt{n}^{\log{n}}\big)$ & 					
		Yes						
		\aftergap
		
		\beforegap
		This paper (sync) &	
		Perfect & 					
		$(1/3-\epsilon)n$ & 		
		No & 						
		No & 						
		$\tilde{O}\big(m + n\sqrt{n}\big)$ & 		
		$\tilde{O}\big(m + n\sqrt{n}\big)$ & 		
		$O\big(d + \mathsf{polylog}(n)\big)$ & 								
		$O(\ell)$ & 			
		Yes                	
		\aftergap		
		
		\beforegap
		This paper (async) &	
		Perfect & 					
		$(1/8-\epsilon)n$ & 		
		Yes & 						
		No & 						
		$\tilde{O}\big(m + n\sqrt{n}\big)$ & 		
		$\tilde{O}\big(m + n\sqrt{n}\big)$ & 		
		$O\big(d + \mathsf{polylog}(n)\big)$ & 								
		$O(\ell)$ & 			
		Yes                	
		\aftergap	
		
		
	\end{tabular}
	
	\footnotesize
	\begin{flushleft}
		\textbf{Parameters:} $n$ is the number of parties; $\ell$ is the size of a field element; $d$ is the depth of the circuit; $\kappa$ is the the security parameter; $\epsilon,\epsilon^{\prime}$ are the positive constants; $Time(\Pi)$ is the worst-case running time of RAM program $\Pi$.
		
		\vspace{1em}
		
		\noindent \textbf{Notes:}\\
		\vspace{0.3em}
		$^\dag$The cost is calculated based on the FHE scheme of~\cite{Brakerski:2012:FHE:2090236.2090262}.\\
		$^\ddag$Assumes a symmetric-key infrastructure. However, unlike the rest, this protocol is secure against an adaptive adversary.
	\end{flushleft}
\end{table*}

\item[Counting Networks.]
The threshold counting problem can be solved in a load-balanced way using \emph{counting networks} that were first introduced by Aspnes~\etal~\cite{Aspnes:1991:CNM:103418.103421}. Counting networks are constructed from simple two-input two-output computing elements called \emph{balancers} connected to one another by wires. A counting network can count any number of inputs even if they arrive at arbitrary times, are distributed unevenly among the input wires, and propagate through the network asynchronously. 

Aspnes~\etal~\cite{Aspnes:1991:CNM:103418.103421} establish an $O(\log^2 n) $ upper bound on the depth complexity of counting networks. Since the latency of counting is dependent on the depth of the network, minimizing this depth has been the goal of many papers in this area. A simple explicit construction of an $O(c^{ \log^*{n}} \log{n})$-depth counting network (for some positive constant $c$), and a randomized construction of an $O(\log n)$-depth counting network that works with high probability are described by Klugerman and Plaxton in~\cite{Klugerman:1992:SCN:129712.129752,Klugerman:1995:SCN:222428}. These constructions use the AKS sorting network~\cite{Ajtai:1983:SN:800061.808726} as a building block.
While this sorting network and the resulting counting networks have $O(\log{n})$ depth and require each party (or gate in their setting) to send $O(\log{n})$ messages, large hidden constants render them impractical.
\end{description}

\section{Preliminaries}\label{sec:prelim}
In this section, we define standard terms, notation, and known building blocks used throughout this paper.

\begin{description}
	\item[Notation.]
	We denote the set of integers $\{1,...,n\}$ by $[n]$. We say an event occurs \emph{with high probability}, if it occurs with probability at least $1-1/n^c$, for some $c>0$ and sufficiently large $n$. A protocol is called \emph{$t$-private} if no coalition of $t$ corrupted parties can learn anything more than what is implied by their private inputs and the protocol output. A protocol is called \emph{$t$-resilient} if no set of $t$ or less parties can influence the correctness of the outputs of the remaining parties.
	
	We also assume that all arithmetic operations in the circuit are carried out over a finite field $\F$. The size of $\F$ depends on the specific function to be computed and is always $\Omega(\log{n})$. All of the messages transmitted by our protocol are logarithmic in $\F$ and $n$.
	
	Let $r$ be a value chosen uniformly at random from $\F$ and $\mask{x} = x + r$, for any $x \in \F$. In this case, we say $x$ \emph{is masked with} $r$ and we refer to $r$ and $\mask{x}$ as the \emph{mask} and the \emph{masked value} respectively.
	
	
	\item[Universal Composability Framework.] When a protocol is executed several times possibly concurrently with other protocols, one requires to ensure this composition preserves the security of the protocol. This is because an adversary attacking several protocols that run concurrently can cause more harm than by attacking a \emph{stand-alone} execution, where only a single instance of one of the protocols is executed. 
	
	One way to ensure this is to show the security of the protocol in the \emph{universal composability (UC)} framework of Canetti~\cite{Canetti:UCSecurity:2001}. A protocol that is secure in the UC framework is called \emph{UC-secure}. We describe this framework in Section~\ref{sec:mainproofs}. 
	 
	
	\item[Verifiable Secret Sharing.]
	An \emph{$(n,t)$-secret sharing} scheme is a protocol in which a dealer who holds a secret value shares it among $n$ parties such that any set of $t<n$ parties cannot gain any information about the secret, but any set of at least
	$t+1$ parties can reconstruct it. An \emph{$(n,t)$-verifiable secret sharing (VSS)} scheme is an \emph{$(n,t)$}-secret sharing scheme with the additional property that after the sharing stage, a dishonest dealer is either disqualified or the honest parties can reconstruct the secret, even if shares sent by dishonest parties are spurious. When we say a set of shares of a secret are \emph{valid}, we mean the secret can be uniquely reconstructed solely from the set of shares distributed among the parties.
	
	In this paper, we use the $(\lceil n/3 \rceil - 1)$-resilient VSS scheme of Ben-Or~\etal~\cite{bgw88} for the synchronous setting and the $(\lceil n/4 \rceil - 1)$-resilient VSS scheme of Ben-Or~\etal~\cite{benor_canetti_goldreich:asynchronous} for the asynchronous setting. When run among $n$ parties, both protocols incur $\poly(n)$ communication cost and $O(1)$ latency. We refer to the sharing stages of these protocols as \alg{VSS-Share} and \alg{AVSS-Share}, and to their reconstruction stages as \alg{VSS-Reconst} and \alg{AVSS-Reconst}, respectively.
	
	\item[Classic MPC.]
	Our main protocols rely on the classic $(\lceil n/3 \rceil - 1)$-resilient MPC protocol of Ben-Or~\etal~\cite{bgw88} for the synchronous setting and the classic $(\lceil n/4 \rceil - 1)$-resilient MPC protocol of Ben-Or~\etal~\cite{benor_canetti_goldreich:asynchronous} for the asynchronous setting. When run among $n$ parties to compute a circuit with $d$ gates, both protocols send $\poly(n)$ bits and incur a latency of $O(d)$. We refer to the former protocol as \alg{CMPC} and to the latter as \alg{ACMPC}.
	
	In this paper, we use the above VSS and classic MPC protocols only among logarithmic-size groups of parties and only for computing logarithmic-size circuits. Thus, the communication overhead per invocation of these protocols will be $\polylog(n)$. 
	
	\item[Byzantine Agreement.]
	In the \emph{Byzantine agreement} problem, each party is initially given an input bit. All honest parties must agree on a bit which coincides with at least one of their input bits. 
	
	When parties only have access to secure pairwise channels, a protocol is required to ensure secure (reliable) broadcast. 
	This guarantees all parties receive the same message even if the broadcaster (dealer) is dishonest and sends different messages to different parties. Every time a broadcast is required in our protocols, we use the Byzantine agreement algorithms of Feldman and Micali~\cite{Feldman:1988:OAB:62212.62225}. We refer to their $(\lceil n/3 \rceil - 1)$-resilient synchronous algorithm as \alg{BA} and to their $(\lceil n/4 \rceil - 1)$-resilient asynchronous algorithm as \alg{ABA}. When all parties participating in a run of a broadcast protocol receive the same message, we say these messages are \emph{consistent}.
\end{description}

\section{Our Protocols} \label{sec:alg}
We now describe our protocols for scalable MPC in large networks. Throughout this section, we consider the network model defined in Section~\ref{sec:model}. We first describe our synchronous protocol, and then adapt this protocol to the asynchronous setting.

We assume that the parties have an arithmetic circuit $C$ computing $f$; the circuit consists of $m$ addition and multiplication gates. For convenience of presentation, we assume each gate has in-degree and out-degree 2.\footnote{Our protocol works, with minor modifications, for gates with arbitrary constant fan-in and fan-out.} For any two gates $x$ and $y$ in $C$, if the output of $x$ is input to $y$, we say that $x$ is a \emph{child} of $y$ and that $y$ is a \emph{parent} of $x$. We assume the gates of $C$ are numbered $1, 2, \dots, m$, where the gate numbered $1$ is the output (root) gate. 

\subsection{Synchronous MPC}
The high-level idea behind our protocols is to first create a sufficient number of quorums and assign to each gate in the circuit one of these quorums. Then, for each party $P_i$ holding an input $x_i \in \F$, $P_i$ secret-shares $x_i$ among all parties in the quorum associated with the $i$-th input gate. We refer to such a quorum as an \emph{input quorum}. 

Next, the protocol evaluates the circuit gate-by-gate starting from input gates. Each gate is jointly evaluated by parties of the quorum associated with this gate over the secret-shared inputs provided by its children. In a similar way, the result of the gate is then used as the input to the computation of the parent gate. Finally, the quorum associated with the root gate, constructs the final result and sends it to all parties via a binary tree of quorums.

This high-level idea relies on solutions to the following main problems.
\begin{description}
	\item[Quorum Building.] Creating a sufficient number of quorums. In Section~\ref{sec:qf}, we describe a randomized protocol called \textsf{Build-Quorums} that achieves this goal with high probability. 
	
	\item[Circuit Evaluation.] Securely evaluating each gate over secret-shared inputs by the parties inside a quorum. In Section~\ref{sec:gateEvaluation}, we describe a protocol called \textsf{Circuit-Eval} that achieves this goal.
	
	\item[Share Renewal.] Sending the result of one quorum to another without revealing any information to any individual party or to any coalition of corrupted parties in both quorums. We solve this as part of our gate evaluation protocol described in Section~\ref{sec:gateEvaluation}.
\end{description}

\smallskip
\mbox{Protocol~\ref{pro:main}} is our main protocol. When we say a party \emph{VSS-shares} (or \emph{secret-shares}) a value $s$ in a quorum $Q$ (or among a set of parties), we mean the party participates as the dealer with input $s$ in the protocol \alg{VSS-Share} with all parties in $Q$ (or in the set of parties).

\begin{algorithm}
	\caption{Synchronous MPC}\label{pro:main}
	\medskip
	\algfont
	\begin{enumerate}
		\item \textbf{Quorum Building.} All parties run \textsf{Build-Quorums} to agree on $n$ good quorums $Q_{1},...,Q_{n}$. The $i$-th gate of $C$ is assigned to $Q_{(i \bmod n)}$, for all $i \in [m]$.
		\item \textbf{Input Commitment.} For all $i \in [n]$, party $P_i$ holding an input value $x_i \in \F$ runs the following steps concurrently:
		
		\smallskip		
		\begin{enumerate}
			\item Pick a uniformly random element $r_i \in \F$, set $\mask{x} = x_i + r_i$, and broadcast  $\mask{x}$ to $Q_i$.
			\item Run \vssSh to secret-share $r_i$ in $Q_i$.
		\end{enumerate}
		
		\item \textbf{Circuit Evaluation.} All parties participate in a run of \ce to securely evaluate $C$.
		
		\item \textbf{Output Reconstruction.} For the output gate $z$, parties in $Q_z$,
		
		\smallskip
		\begin{enumerate}
			\item Run \vssRec to reconstruct $r_z$ from its shares.
			\item Set the circuit output message: $y \gets \mask{y}_z - r_z$.
			\item Send $y$ to all parties in the $Q_2$ and $Q_3$.			
		\end{enumerate}
		
		\item \textbf{Output Propagation.} For every $i \in \{2,...,n\}$, parties in $Q_i$ perform the following steps:
		\smallskip
		\begin{enumerate}
			\item Receive $y$ from the $Q_{\lfloor i/2\rfloor}$.
			\item Send $y$ to all parties in $Q_{2i}$ and	$Q_{2i+1}$.
		\end{enumerate}		
	\end{enumerate}
\end{algorithm}

The protocol starts by running \textsf{Build-Quorums} to create $n$ quorums $Q_1,...,Q_n$. Then, it assigns the gates of $C$ to these quorums in the following way. The output gate of $C$ is assigned to $Q_1$; then, every gate in $C$ numbered $i$ (other than the output gate) is assigned to $Q_{(i \bmod n)}$. For each gate $u \in C$, we let $Q_u$ denote the quorum associated with $u$, $y_u$ denote the output of $u$, $r_u$ be a random element from $\F$, and $\mask{y}_v$ denote the masked output of $u$, where $\mask{y}_u = y_u + r_u$.

\begin{figure*}[t]
	\begin{center}
		\includegraphics[scale=0.35]{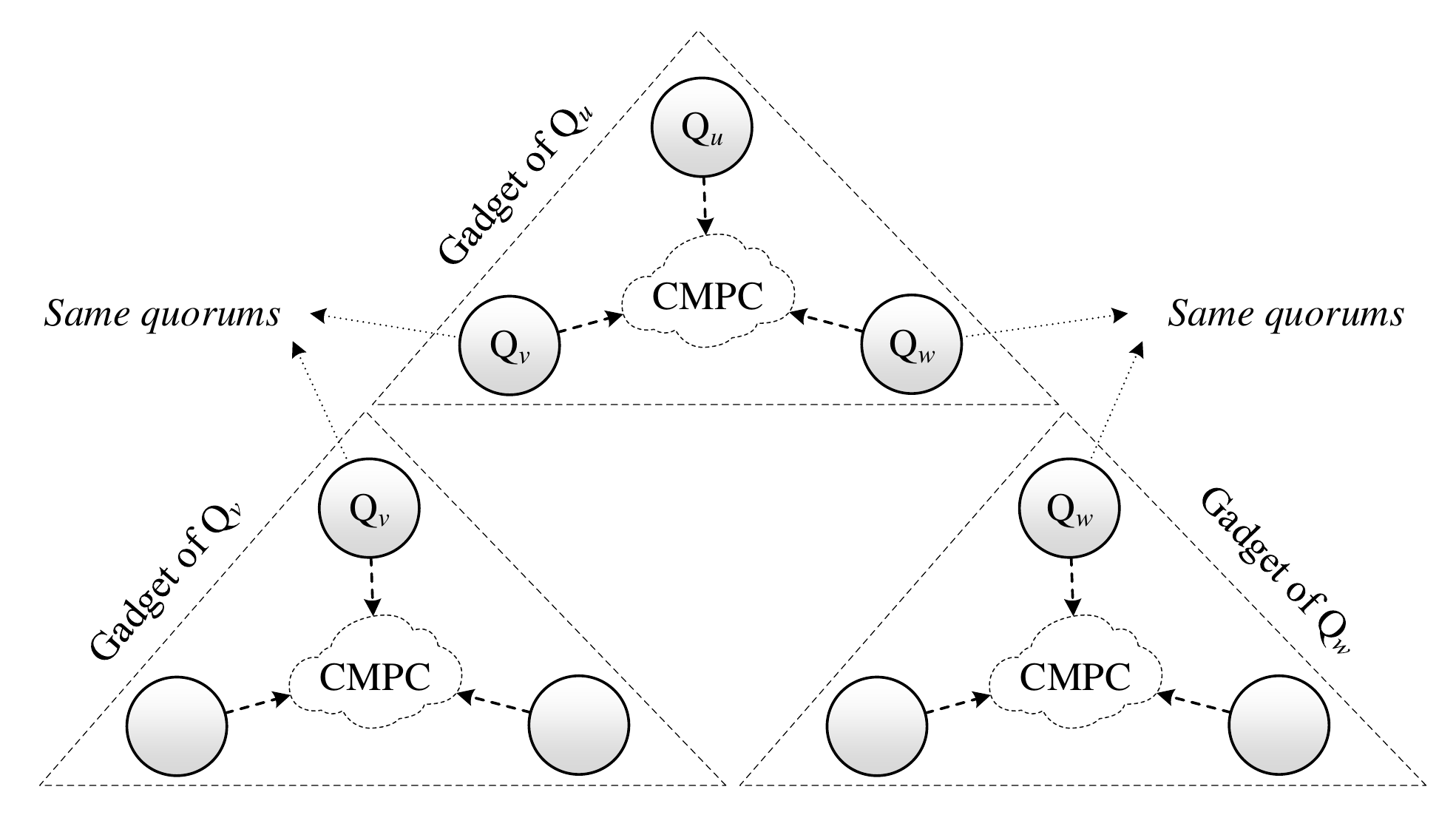}
	\end{center}
	\caption{The gate gadgets for gate $u$ and its left and right children}
	\label{fig:gadget}
\end{figure*}

\subsubsection{Input Commitment}
Let $Q_i$ be the quorum associated with party $P_i$ who holds input $x_i$. At the start of our protocol, $P_i$ samples a value $r_i$ uniformly at random from $\F$, sets $\mask{x} = x_i + r_i$, and broadcasts $\mask{x}$ to all parties in $Q_i$. Next, $P_i$ runs \alg{VSS-Share} to secret-share $r_i$ among all parties in $Q_i$.

\subsubsection{Circuit Evaluation} \label{sec:gateEvaluation}
The main idea for reducing the amount of communication required in evaluating the circuit is  quorum-based gate evaluation. If each party participates in the computation of the whole circuit, it must communicate with all other parties. Instead, in quorum-based gate evaluation, each gate of the circuit is computed by a \emph{gate gadget}. A gate gadget (see Figure~\ref{fig:gadget}) consists of three quorums: two \emph{input quorums} and one \emph{output quorum}. Input quorums are associated with the gate's children which serve inputs to the gate. The output quorum is associated with the gate itself and is responsible for creating a shared random mask and maintaining the output of the quorum for later use in the circuit. As depicted in Figure~\ref{fig:gadget}, these gate gadgets connect to form the entire circuit. In particular, for any gate $u$, the output quorum of $u$'s gate gadget is the input quorum of the gate gadget for all of $u$'s parents.

The parties in each gate gadget run \hw among themselves to compute the gate operation. To ensure privacy is preserved, each gate gadget maintains the invariant that the value computed by the gadget is the value that the corresponding gate in the original circuit would compute, masked by a uniformly random element of the field. This random element is not known to any individual party. Instead, shares of it are held by the members of the output quorum. Thus, the output quorum can participate as an input quorum for the evaluation of any parent gate and provide both the masked version of the inputs and shares of the mask.
The gate gadget computation is performed in the same way for all gates in the circuit until the final output of the whole circuit is computed. After the input commitment step, for each input gate $u$, parties in $Q_u$ know the masked input $\mask{y}_u$, and each has a share of the mask $r_u$.
\begin{figure*}[t]
	\begin{center}
		\includegraphics[scale=0.34]{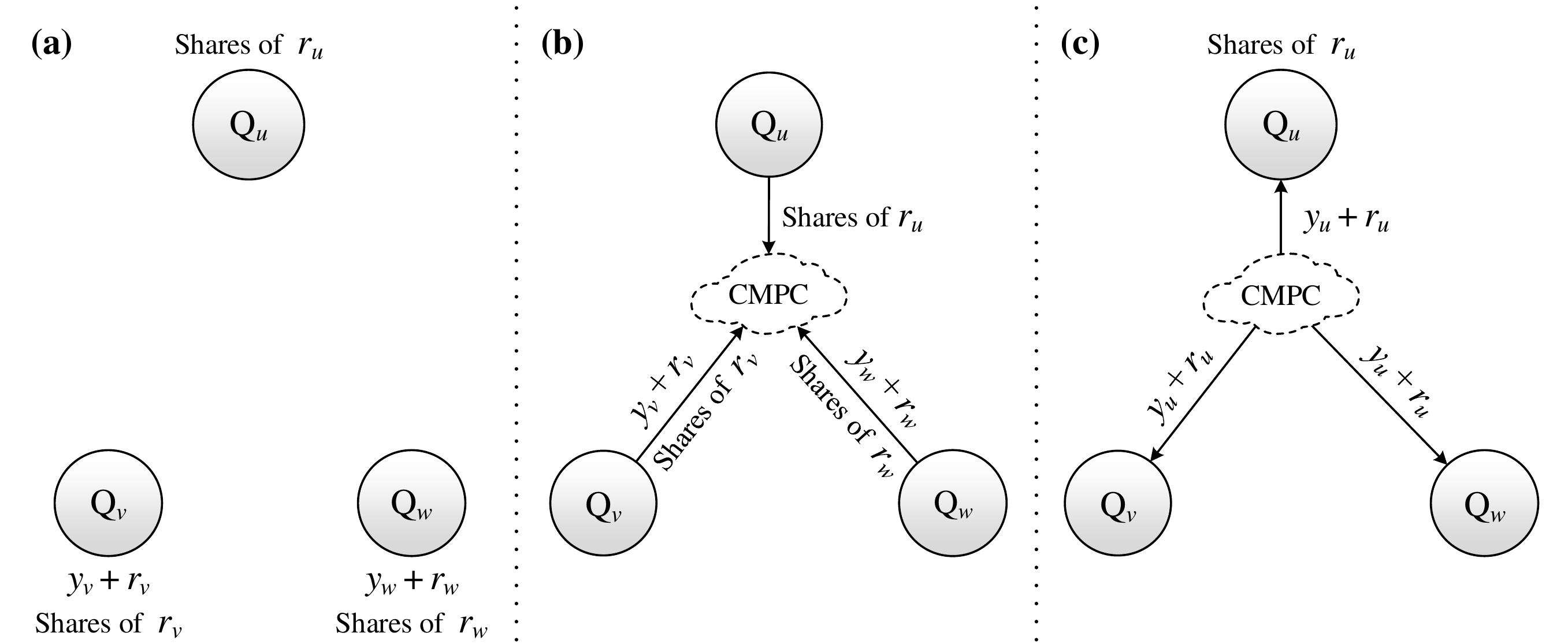}
	\end{center}
	\caption{Evaluation of gate $u$: (a) generating $r_u$, (b) providing inputs to \alg{CMPC}, (c) receiving the masked outputs }
	\label{f:quorums}
\end{figure*}

The first step of the circuit evaluation is to generate shares of uniformly random field elements for all gates. If a party is in a quorum at gate $u$, it generates shares of $r_u$, a uniformly random field element, by participating in the \alg{Gen-Rand} protocol. These shares are needed as inputs to the subsequent run of \hw.

\begin{algorithm}
	\caption{\ce}\label{pro:ce}
	\medskip
	\algfont
	\textit{Goal.} Given a circuit $C$, the protocol securely evaluates $C$.
	
	\smallskip
	For every gate $u \in C$ with children $v,w \in C$, parties in $Q_u$, $Q_v$, and $Q_w$ perform the following steps to compute the gate functionality:
	\begin{enumerate}
		\item \textbf{Mask Generation.} Parties in $Q_u$ run \alg{Gen-Rand} to jointly generate a secret-shared random value \mbox{$r_u \in \F$}.
		
		\item \textbf{MPC in Quorums.} The following parties participate in a run of \hw with their corresponding inputs:
		\begin{itemize}
			\item Every party in $Q_u$ with his share of $r_u$.
			\item Every party in $Q_v$ with his input \\$\left(\mask{y}_v,~\text{his share of }r_v\right)$.
			\item Every party in $Q_w$ with his input \\$\left(\mask{y}_w,~\text{his share of }r_w\right)$.			
		\end{itemize}
		
	\end{enumerate}
\end{algorithm}

Next, parties form the gadget for each gate $u$ to evaluate the functionality of the gate using~\ce. Let $v$ and $w$ be the left and right children of $u$ respectively. The gate evaluation process is shown in Figure~\ref{f:quorums}. The values $y_v$ and $y_w$ are the inputs to $u$, and $y_u$ is the its output as it would be computed by a trusted party. Each party in $Q_u$ has a share of the random element $r_u$ via \alg{Gen-Rand}. Every party in $Q_v$ has the masked value $y_v + r_v$ and a share of $r_v$ (respectively for $Q_w$). 

As shown in Part~(b) of Figure~\ref{f:quorums}, all parties in the three quorums participate in a run of \hw, using their inputs, in order to compute $\mask{y}_u = y_u + r_u$. Part~(c) of the figure shows the output of the gate evaluation after participating in \hw. Each party in $Q_u$ now learns $\mask{y}_u$ as well a share of $r_u$. Therefore, parties in $Q_u$ now have the input required for performing the computation of parents of $u$ (if any). Note that both $y_u$ and $r_u$ remain unknown to any individual.

The gate evaluation is performed for all gates in $C$ starting from the bottom to the top. The output of the quorum associated with the output gate in $C$ is the output of the entire algorithm. This quorum will unmask the output via the output reconstruction step. The last step of the algorithm is to send this output to all parties. We do this via a complete binary tree of quorums, rooted at the output quorum.
\begin{algorithm}
	\caption{\alg{Gen-Rand}} \label{pro:GenRand}
	\medskip
	\algfont
	\textit{Goal.} A set of parties $P_1,...,P_N$ in a quorum want to agree on a secret-shared value $r$ chosen uniformly at random from $\F$.
	
	\begin{enumerate}
		\item For all $i\in[N]$, party $P_i$ chooses $\rho_i \in \mathbb{F}$ uniformly at random and VSS-shares it among all $N$ parties.
		\item For every $j\in[N]$, let $N^\prime$ be the number of shares $P_j$ receives from the previous step, and $\rho_{1j},...,\rho_{N^\prime j}$ be these shares. $P_j$ computes $r_j=\sum_{k=1}^{N^\prime}\rho_{kj}$.	
	\end{enumerate}
\end{algorithm}

\subsubsection{Implementing the Gate Circuit} 
For every gate $u \in C$, the \alg{Circuit-Eval} protocol requires a circuit (as we denote by $C_u$) for unmasking the masked inputs $\mask{y}_v$ and $\mask{y}_w$, computing $u$'s functionality $f_u$ over the unmasked inputs, and masking the output with the gate's random value $r_u$. This circuit is securely evaluated using the \alg{CMPC} protocol by the quorum associated with~$u$. 

\begin{figure*}[t]
	\begin{center}
		\includegraphics[scale=0.68,trim={0 3cm 0 3.5cm},clip]{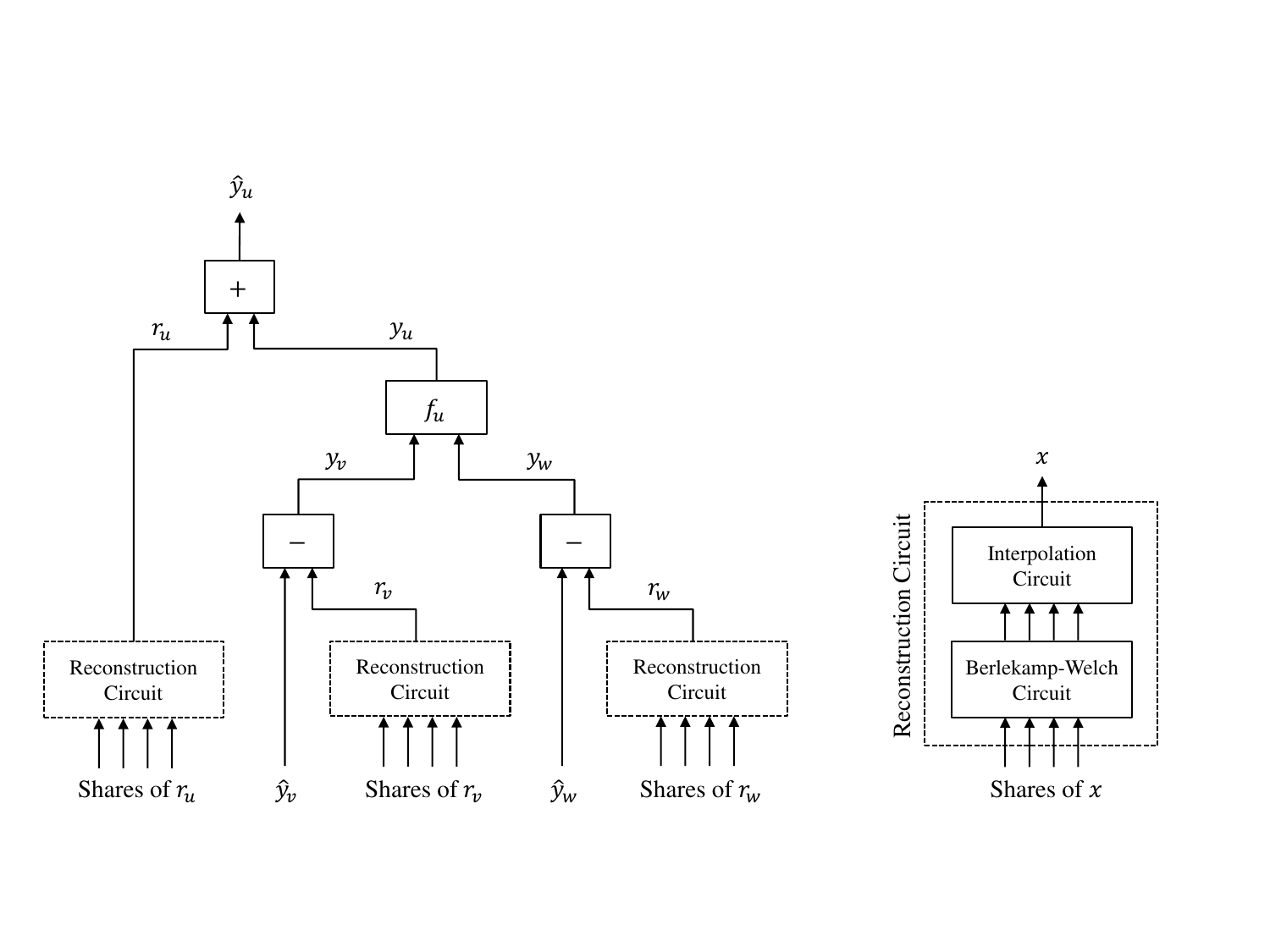}
	\end{center}
	\caption{Circuit of gate $u$}
	\label{f:gateCircuit}
\end{figure*}

For unmasking an input, $C_u$ requires a \emph{reconstruction circuit}, which given a set of shares, outputs the corresponding secret. Since dishonest parties may send spurious shares, the circuit implements the error-correcting algorithm of Berlekamp and Welch~\cite{Berlekamp:Welch:1986} to fix such corruptions. Then, the resulting shares are given to an \emph{interpolation circuit} which implements a simple polynomial interpolation. Figure~\ref{f:gateCircuit} depicts the circuit for gate~$u$.

We now briefly describe the error correcting algorithm of Berlekamp and Welch~\cite{Berlekamp:Welch:1986}. Let $\mathbb{F}_{p}$ denote a finite field of prime order $p$, and $S=\{(x_{1},y_{1})\:|\:x_{i},y_{i}\in\mathbb{F}_{p}\}_{i=1}^{\eta}$ be a set of $\eta$ points, where $\eta-\varepsilon$ of them are on a polynomial $y=P(x)$ of degree $\tau$, and the rest $\varepsilon<(\eta-\tau+1)/2$ points are erroneous. Given the set of points $S$, the goal is to
find the polynomial $P(x)$. The algorithm proceeds as follows. Consider two polynomials $E(x)=e_{0}+e_{1}x+...+e_{\varepsilon}x^{\varepsilon}$ of degree $\varepsilon$, and $Q(x)=q_{0}+q_{1}x+...+q_{k}x^{k}$ of degree $k\leq\varepsilon+\tau-1$ such that $y_{i}E(x_{i})=Q(x_{i})$
for all $i\in[\eta]$. This defines a system of $\eta$ linear equations with $\varepsilon+k=\eta$ variables $e_{0},...,e_{\varepsilon},q_{0},...,q_{k}$ that can be solved efficiently using Gaussian elimination technique
to get the coefficients of $E(x)$ and $Q(x)$. Finally, calculate $P(x)=Q(x)/E(x)$.

Since the Gaussian elimination algorithm over finite fields has $O(n^3)$ arithmetic complexity~\cite{1988linear}, the corresponding circuit has at most $O(n^3)$ levels. Since the interpolation circuit consists of at most $O(n^2)$ arithmetic operations (using the Lagrange's method~\cite{Abramowitz:1974:HMF:1098650}), the overall depth of the reconstruction circuit will be $O(n^3)$.

\subsection{Asynchronous MPC}
We now adapt our synchronous protocol to the asynchronous communication model. We do this by modifying the following parts of Protocol~\ref{pro:main}:

\begin{enumerate}
	\item We replace the synchronous subprotocols \alg{VSS-Share}, \alg{VSS-Reconst}, and \alg{CMPC} with their corresponding asynchronous versions \alg{AVSS-Share}, \alg{AVSS-Reconst}, and \alg{ACMPC}, respectively. In Section~\ref{sec:qf}, we describe a technique for adapting \alg{Build-Quorums} to the asynchronous setting.
	
	\item At the end of the Input Commitment stage, the protocol should wait for at least $n-t$ inputs before proceeding to the Circuit Evaluation stage. To this end, we introduce a new subprotocol called \alg{Wait-For-Inputs} and invoke it right after step~(b) of the Input Commitment stage. This protocol is described in Section~\ref{sec:waitforinputs}.
	
	\item Although the protocol \alg{ACMPC} terminates with probability one, its actual running time (\ie, the number of rounds until it terminates) is a random variable with expected value $O(D\log{N})$, where $N$ is the number of parties participating in the MPC and $D$ is the circuit depth~\cite{benor_canetti_goldreich:asynchronous}. Since we run $m$ instances of \alg{ACMPC} (one for each gate of $C$), we need a method that allows us to bound the running time of each gate, and thus to bound the expected running time of our asynchronous MPC protocol. We describe a simple method for achieving this in Section~\ref{sec:runtime}.
	
	\item In the second step of \alg{Gen-Rand} (Protocol~\ref{pro:GenRand}), each party may receive less than $N$ shares. This is because  
\end{enumerate}

\subsubsection{Implementing \alg{Wait-For-Inputs}} \label{sec:waitforinputs}
The protocol \alg{Wait-For-Inputs} counts the number of inputs that are successfully received by their corresponding input quorums. This can be achieved using a solution to the threshold counting problem: Count the number of inputs successfully received by each input quorum and return once this number becomes greater than or equal to $n-t$. As a result of returning from \alg{Wait-For-Inputs}, the main protocol resumes and starts the circuit evaluation procedure.

In Section~\ref{sec:taucount}, we provide a solution to the threshold counting problem. We refer to this protocol as \alg{Thresh-Count}. This protocol creates a distributed tree structure called the \emph{count tree} which is known to all parties and determines how the parties communicate with each other to count of the number of inputs.

Protocol~\ref{pro:waitForInputs} implements \textsf{Wait-For-Inputs} using our \alg{Thresh-Count} algorithm. In \alg{Wait-For-Inputs}, the role of each party in \alg{Thresh-Count} (\ie, each node in the count tree) is played by a quorum of parties. Once \alg{Thresh-Count} terminates, the parties in each input quorum decide whether or not the corresponding inputs are part of the computation.

When run among quorums, \alg{Thresh-Count} requires the quorums to communicate with each other. We say a quorum $Q$ sends a message $M$ to quorum $Q^\prime$, when every (honest) party in $Q$ sends $M$ to every party in $Q^\prime$. A party in $Q^\prime$ is said to have received $M$ from $Q$ if it receives $M$ from at least $\fgoodq$ of the parties in $Q$. When we say a party \emph{broadcasts} a message $M$ to a quorum $Q$, we mean the party sends $M$ to every party in $Q$, and then, all parties in $Q$ run \ba over their messages to ensure they all hold the same message. 

\begin{algorithm}
	\caption{\alg{Wait-For-Inputs}}\label{pro:waitForInputs}
	\medskip
	\algfont
	\textit{Goal.} For every input quorum $Q$, all parties in a quorum $Q$ wait until $n-t$ inputs are received by the input quorums. For each party $P_i \in Q$, $P_i$ is initially
	holding two values $\mask{x}$ and $r_i$, the $i$-th share of a random value $r$.
	
	\medskip
	Each party $P_i \in Q$ does the following:
	\smallskip
	\begin{enumerate}
		\item Run \alg{Thresh-Count} with flag bit $b_i$ initially set to zero.
		
		\item If $\mask{x}$ and $r_i$'s are consistent and valid (based on the Byzantine agreement protocol and the verification stage of \alg{AVSS-Share} respectively), set $b_i \gets 1$ in Step $2(a)$ of \alg{Thresh-Count}.
		
		\item Upon receiving \done from the parent quorum, run \alg{ACMPC} using $b_i$ as the input. $d \gets \textsf{True}$ if a $5/8$-fraction of the parties in $Q$ have their $b_i$'s set to one. Otherwise, $d \gets \textsf{False}$.\label{ln:waitThreshCMPC}
		
		\item If $d = \textsf{False}$, then $\mask{x} \gets \textsf{Default}$ and $r_i \gets 0$.
	\end{enumerate}	
\end{algorithm}

\subsubsection{Bounding the Expected Running Time} \label{sec:runtime}
Consider $N$ parties in a quorum who want to jointly compute a circuit of depth $D$ using the protocol \alg{ACMPC}. Let $X$ denote the random variable corresponding to the number of rounds until an instance of \alg{ACMPC} terminates. From~\cite{benor_canetti_goldreich:asynchronous}, we have \[\centering \mathbf{E}[X] = O(D\log{N}).\] Instead of running only one instance of \alg{ACMPC}, we run $O(\log{N})$ instances sequentially each for $2\mathbf{E}[X]$ rounds. The output corresponding to the first instance that terminates will be returned as the output of the gate. Using the Markov's inequality, \[\Pr(X \geq 2\mathbf {E}[X]) \leq 1/2.\] 

In each gate of $C$, each party also participates in a run of \alg{Gen-Rand} which itself calls \alg{AVSS-Share}. Similar to \alg{ACMPC}, for each instance of \alg{AVSS-Share}, we run $O(\log{N})$ instances sequentially each for $2\mathbf{E}[X]$ rounds. The sharing corresponding to the first instance that terminates will be accepted by the parties.

Since $O(\log{N})$ instances of \alg{ACMPC} and \alg{AVSS-Share} are executed in each gate, the computation of the gate correctly terminates after at most \[2\mathbf{E}[X]\log{N} = O(D\log^2{N})\] rounds with high probability.
Since $C$ has $m = \mathsf{poly}(n)$ gates, using union bound over all gates of $C$, our MPC algorithm correctly terminates with high probability. Finally, since $C$ has depth $d$, the expected running time of our asynchronous MPC protocol is $O(Dd\log^2{N})$. In Section~\ref{sec:gateEvaluation}, we argued that the circuit computed by \alg{Circuit-Eval} has depth $D = \mathsf{polylog}(n)$. Thus, the expected running time of our protocol is $O(d \polylog(n))$.

\subsection{Remarks}
\label{sec:dis}
As described in the introduction, the goal of MPC is to simulate a trusted third party in the computation of the circuit, and then send back the computation result to the parties. Let $S$ denote the set of parties from whom input is received by the (simulated) trusted party. Recall that $|S| \ge n-t$.\footnote{We allow $|S| > n-t$ because the adversary is not limited to delivering one message at a time; two or more messages may be received simultaneously.}
Thus, for an arbitrary $S$, a description of $S$ requires $\Omega(n)$ bits, and cannot be sent back to the parties using only a scalable amount of communication. Therefore, we relax the standard requirement that $S$ be sent back to the parties. Instead, we require that at the end of the protocol each honest party learns the output of $f$; whether or not their own input was included in $S$; and the \emph{size} of $S$.

Also note that although we have not explicitly included this in the input commitment step, it is very easy for the parties to compute the size of the computation set $S$. Once each input quorum $Q_i$ has performed the third step of \alg{Wait-For-Inputs} and has agreed on the flag $b_i = 1$, they can simply use an addition circuit to add these bits together, and then disperse the result. This is an MPC, all of whose inputs are held by honest parties, since each input flag $b_i$ is jointly held by the entire quorum $Q_i$, and all the quorums are good. Thus, the computation can afford to wait for all $n$ inputs and computes the correct sum.

In our both protocols, it may be the case that a party $P$ participates more than one time in the quorums performing a single instance of the classic MPC. In such a case, we allow $P$ to play the role of more than one different parties in \alg{CMPC} and \alg{ACMPC}, one for each quorum to which $P$ belongs. This ensures that the fraction of corrupted parties in any instance of the classic MPC is always less than $1/3$ for the synchronous case and $1/4$ for the asynchronous case. We stress that \alg{CMPC} and \alg{ACMPC} both maintain privacy guarantees even in the face of gossiping coalitions of constant size. Thus, each party will learn no information beyond the output and its own inputs after running these protocols.

\section{Proof of Theorem~\ref{thm:main}} \label{sec:mainproofs}
We first describe the UC framework in Section~\ref{sec:ucframework}, and then give a sketch of our proof in Section~\ref{sec:ucproof}. We prove the UC-security of Protocol~\ref{pro:main} in sections~\ref{sec:proof-input} to \ref{sec:proof-output}. Finally, we calculate the resource costs of this protocol in Section~\ref{sec:costAnalysis}.

\subsection{The UC Framework} \label{sec:ucframework}
The UC framework is based on the \emph{simulation paradigm}~\cite{Goldreich:2000:FCB:519078}, where the protocol is considered in two models: \emph{ideal} and \emph{real}. In the ideal model, the parties send their inputs to a trusted party who computes the function and sends the outputs to the parties. We refer to the algorithm run by the trusted party in the ideal model as the \emph{functionality} of the protocol. In the real model, parties run the actual protocol that assumes no trusted party. We refer to a run of the protocol in one of these models as the \emph{execution} of the protocol in that model.

A protocol $\mathcal{P}$ securely computes a functionality \func{$\mathcal{P}$} if for every adversary $\mathcal{A}$ in the real model, there exists an adversary $\mathcal{S}$ in the ideal model, such that the result of a real execution of $\mathcal{P}$ with $\mathcal{A}$ is indistinguishable from the result of an ideal execution with $\mathcal{S}$. The adversary in the ideal model, $\mathcal{S}$, is called the \emph{simulator}.


The simulation paradigm provides security only in the stand-alone model. To prove security under composition, the UC framework introduces an adversarial entity called the \emph{environment}, denoted by $\mathcal{Z}$, who generates the inputs to all parties, reads all outputs, and interacts with the adversary in an arbitrary way throughout the computation. The environment also chooses inputs for the honest parties and gets their outputs when the protocol is finished.

A protocol is said to \emph{UC-securely} compute an ideal functionality if for any adversary $\mathcal{A}$ that interacts with the protocol there exists a simulator $\mathcal{S}$ such that no environment $\mathcal{Z}$ can tell whether it is interacting with a run of the protocol and $\mathcal{A}$, or with a run of the ideal model and $\mathcal{S}$.

Now, consider a protocol $\mathcal{P}$ that has calls to $\ell$ subprotocols $\mathcal{P}_1,...,\mathcal{P}_\ell$ which are already proved to be UC-secure. To facilitate the security proof of $\mathcal{P}$, we can make use of the \emph{hybrid model}, where the subprotocols are assumed to be ideally computed by a trusted third-party. In other words, we replace each call to a subprotocol with a call to its corresponding functionality. This hybrid model is usually called the \emph{$(\mathcal{P}_1,...,\mathcal{P}_\ell)$-hybrid} model. We say $\mathcal{P}$ is \emph{UC-secure in the hybrid model} if $\mathcal{P}$ in the hybrid model is indistinguishable by the adversary from $\mathcal{P}$ in the ideal model.
The \emph{modular composition theorem}~\cite{Canetti:2000:JC:0933-2790} states that if $\mathcal{P}_1,...,\mathcal{P}_\ell$ are all UC-secure, and $\mathcal{P}$ is UC-secure in the hybrid model, then $\mathcal{P}$ is UC-secure in the real model.

\subsection{Proof Sketch} \label{sec:ucproof}
Before proceeding to the proof, we remark that the error probability in Theorem~\ref{thm:main} comes entirely from the possibility that \alg{Build-Quorums} or the threshold counting procedure may fail to output correct results. All other components of our protocol are deterministic and thus have no error probability. We also assume that, at the beginning of our MPC protocol, the parties have already agreed on $n$ good quorums, and the threshold counting procedure is performed successfully.\footnote{For simplicity, we assume the primitive \alg{Build-Quorums} is run only once, and it does not run concurrently with other protocols.}

As in~\cite{Goldreich:2004:FCV:975541}, we refer to the security in the presence of a malicious adversary controlling $t$ parties \emph{$t$-security}. For every gate $u \in C$, let $I_u$ denote the set of the corrupted parties in the quorum associated with $u$. Also, let $I$ denote the set of all corrupted parties, where $|I|<t$. 

Our goal is to prove the UC-security of Protocol~\ref{pro:main}.  To do this, we must show two steps.  Step 1) is to show that each of our subprotocols are UC-secure.  Step 2) is to show that our protocol is UC-secure in the hybrid model.  Once we show these two steps, then by the modular composition theorem, we conclude that our protocol is UC-secure in the real model.

In Lemma~\ref{lem:step2}, we show Step 2, that the adversary can not distinguish the execution of the hybrid model from the ideal model.

We next describe our approach to Step 1, which is more challenging.  For this step, we make use of a theorem that will help us show that our subprotocols are UC-secure. Kushilevitz~\etal~\cite{Kushilevitz:2010:ISP:1958016.1958032} show Theorem~\ref{thm:ucperfect}. This theorem targets perfectly-secure protocols that are shown secure using a straight-line black-box simulator. A black-box simulator is a simulator that is given only oracle access to the adversary (see~\cite{Goldreich:2000:FCB:519078} Section 4.5 for a detailed definition). Such a simulator is straight-line if it interacts with the adversary in the same way as real parties, meaning that it proceeds round by round without ever going back.

\begin{theorem}[\cite{Kushilevitz:2010:ISP:1958016.1958032}] \label{thm:ucperfect}
	Every protocol that is perfectly-secure in the stand-alone model and has a straight-line black-box simulator is UC-secure.
\end{theorem}

 We first define the ideal functionalities shown in Table~\ref{tbl:functionalities} that correspond to the subprotocols used in Protocol~\ref{pro:main}. We then prove that Protocol~\ref{pro:main} is $t$-secure in the (\func{BA}, \func{VSS-Share}, \func{VSS-Reconst}, \func{CMPC}, \func{Gen-Rand}, \func{Input}, \func{Circuit-Eval}, \func{Output})-hybrid model. Finally, we use Theorem~\ref{thm:ucperfect} to infer the UC-security of Protocol~\ref{pro:main}.

In order to prove the $t$-security of Protocol~\ref{pro:main} in the hybrid model, we first show that all of our subprotocols are UC-secure. Similar to the above approach, we first prove $t$-security of every subprotocol in its corresponding hybrid model using a straight-line black-box simulator, and then use Theorem~\ref{thm:ucperfect} to infer its UC-security.

\begin{table}
	\caption{Ideal functionalities}
	\label{tbl:functionalities}
	\centering
	\begin{tabular}{|c|l|}
		\hline Functionality  & Implemented by \\ 
		\hline \func{BA} & Protocol \textsf{BA} \\ 
		\func{VSS-Share} & Protocol \textsf{VSS-Share}  \\ 
		\func{VSS-Reconst} & Protocol \textsf{VSS-Reconst}  \\ 
		\func{CMPC} & Protocol \textsf{CMPC} \\ 
		\func{Gen-Rand} & Protocol~\textsf{Gen-Rand} \\ 
		\func{Input} & Input Commitment stage of Protocol~\ref{pro:main} \\
		\func{Circuit-Eval} & Protocol~\textsf{Circuit-Eval} \\
		\func{Output} & Output Reconstruction and Output \\ & Propagation stages of Protocol~\ref{pro:main} \\
		\hline 
	\end{tabular}
\end{table}

To prove the $t$-security of a protocol $\Pi$, we describe a simulator $\mathcal{S}_\Pi$ that simulates the real protocol execution by running a copy of $\Pi$ in the ideal model. For each call to a secure subprotocol $\pi$, the simulator calls the corresponding ideal functionality $F_\pi$. A \emph{view} of a corrupted party from execution of a protocol is defined as the set of all messages it receives during the execution of that protocol. At every stage of the simulation process, $\mathcal{S}_\Pi$ adds the messages received by every corrupted party in that stage to its view of the simulation. This is achieved by running a copy of $\Pi$ for each corrupted party with its actual input as well as by running a copy of $\Pi$ for each honest party with a dummy input.\footnote{$\mathcal{S}_\Pi$ learns neither the actual inputs nor the actual outputs of the honest parties.} The view of the adversary is then defined as the combined view of all corrupted parties.

\subsection{Security of Input Commitment} \label{sec:proof-input}
Before proceeding to the proof of security for Input Commitment stage, we show the following auxiliary lemma.

\begin{lemma}\label{lem:qcomm}
	If a quorum $Q$ sends to a quorum $Q^\prime$ a message $M$, it is eventually received by all honest parties in $Q^\prime$.
\end{lemma}
\begin{proof} 
	Recall that when $Q$ sends $M$ to $Q^\prime$, every honest party in $Q$ sends $M$ to all parties in $Q^\prime$. A party in $Q^\prime$ considers itself to have received the message $M$ from $Q$ if it receives $M$ from at least \fgoodq of the parties in $Q$. Since $n$ quorums have successfully been formed, more than \fgoodq of the parties in each quorum are honest.
	In particular, this is true for $Q$. Thus, at least \fgoodq of the members of $Q$ send $M$ to each member of $Q^\prime$.
	Since the adversary must eventually deliver all the messages that have been sent, albeit with arbitrary delays, it follows that eventually each honest party in $Q^\prime$ receives $M$ from at least \fgoodq of the members of $Q$.
	\qed
\end{proof}

We now proceed to the proof of the Input Commitment stage. The ideal functionality, \func{Input}, is given in Protocol~\ref{fun:Input}. This functionality creates a set $S$ containing the index of the parties whose inputs have been accepted (as defined in Step 1 of Protocol~\ref{fun:Input}) by the protocol to be used for the computation. If a party's input is not in $S$, then the functionality sets this input to the default value. Next, the functionality sends each masked input $\mask{x_i}$ to quorum $Q_i$ and secret-shares the mask $r_i$ in $Q_i$. In Lemma~\ref{lem:inputuc}, we show the Input Commitment stage in Protocol~\ref{pro:main} correctly implements this functionality. Thus, the parties in $Q_i$ eventually either have received consistent VSS-shares of $x_i$ and have agreed on $\mask{x_i} = x_i+r_i$ as well as on $i$ being in $S$ or they have agreed that $i \notin S$ and have set these values to the predefined value and $r_v$ and all its shares to 0.
We say that a quorum has come to agreement on $X$ if all honest parties in the quorum agree on $X$.

\begin{algorithm}
	\algfont
	\caption{\func{Input}}
	\label{fun:Input}
	
	\textit{Goal.} The functionality guarantees valid inputs are received by at least $n-t$ input quorums. Then, the functionality notifies all input quorums to proceed to the next stage of the protocol with either a valid input or a default input.
	\medskip
	
	\textit{Functionality:}
	\begin{enumerate}
		\item Wait to receive at least $n-t$ valid inputs from the set of all $n$ parties. For every such input $x_i$, the functionality receives $\mask{x}_i = x_i + r_i$ and $r_i$ from party $P_i$ where $i\in[n]$. Let $S$ denote the set of parties whose inputs have been accepted (Note that if $P_i \in S$, then $\mask{y}_i$ and $r_i$ are valid).
		
		\item  If $P_i \notin S$, then define $\mask{y}_i =0$ and $r_i = 0$.
		
		\item Broadcast $\mbox{\done}$ and $y_i{\leftarrow}{x_i+r_i}$ to $Q_i$ and run \func{VSS-Share} to secret-share $r_i$ in $Q_i$.	
	\end{enumerate}
\end{algorithm} 

\begin{lemma} \label{lem:inputuc}
	The Input Commitment stage of Protocol~\ref{pro:main} is UC-secure.
\end{lemma}
\begin{proof}
	First, we show that corrupted parties cannot do anything but choose their input as they wish; thus, the Input Commitment stage correctly computes \func{Input}. This means that all honest parties receive the $\mbox{\done}$ message. Moreover, there exists a set $S$ such that for every $i \in [n]$, the following statements hold:
	\begin{enumerate}
		\item All parties in $Q_i$ eventually agree whether $i \in S$ or not.
		
		\item At least $n-t$ input quorums agree that their corresponding party's index is in $S$.
		
		\item All parties in $Q_i$ agree that party $i \in S$ if and only if they collectively hold enough shares to reconstruct $P_i$'s input. If all parties in $Q_i$ agree that $i \in S$, then party $P_i$'s input will be used in the computation. Otherwise, the default value will be used instead.
	\end{enumerate}

	First, since there are $n-t$ honest parties, at least $n-t$ valid inputs are eventually sent to \alg{Thresh-Count}. Based on Theorem~\ref{thm:taucount}, all parties will be notified when $n-t$ inputs are received.
	
	Each party in $Q_i$ has set its flag bit to either 1 or 0 depending on whether it has received a valid input share from $P_i$. Let $q = |Q_i|$. Upon receiving the \done message, the parties in $Q_i$ run the third step of \alg{Wait-For-Inputs} to decide whether at least \threshq of them have set their flag bit to 1. If they have, they assume $i \in S$.
	
	If $i \in S$, then at least \fgoodqq of the parties in $Q_i$ have received input shares from $P_i$ before they received the \done message. Of these, more than $\frac{3q}{4}$ parties in $Q_i$ are honest and have set their flag bit to 1. Since \hw in Line~\ref{ln:waitThreshCMPC}  starts even if as many as \fbadq inputs are missing, the parties in $Q_i$ will correctly decide that at least \threshq flag bits among them are set to 1. Thus, the parties in $Q_i$ all agree that $i \in S$.
	If $i \notin S$, then \hw in Line~\ref{ln:waitThreshCMPC} has determined that less than \threshq flag bits are set to 1. Since $Q_i$ contains less than \fbadq corrupted parties, more than $q/2$ parties set their flags to 0 and the parties in $Q_i$ all agree that $i \notin S$. As a result, at least $n-t$ input quorums agree that their corresponding inputs are in $S$, and hence $|S| \ge n-t$.

		
	We prove the $t$-security of the Input Commitment stage in the (\func{VSS-Share}, \func{VSS-Reconst},\func{CMPC})-hybrid model which is similar to the Input Commitment stage of Protocol~\ref{pro:main} except that every call to its subprotocols is replaced with a call to their corresponding functionality. We define the corresponding simulator \simu{Input} in Protocol~\ref{pro:InputSim}.

	\begin{algorithm}
		\algfont
		\caption{\simu{Input}}
		\label{pro:InputSim}
		
		For every $i \in [n]$, party $P_i$ holds an input $x_i \in \F$. Associated with this input, we consider a quorum $Q_i$. Let $I_i$ denote the set of corrupted parties in $Q_i$, and let $I$ denote the set of all corrupted parties among $P_1,...,P_n$.\medskip
		
		\textit{Inputs.} $\{r_i\}_{i\in [n]}$, and $\{\mask{x}_i\} _{i \in [n]}$ from parties in $I$ (set of all corrupted parties). \medskip
		
		\textit{Simulation:}
		
		\begin{enumerate}
				\item For every $i \in [n]$, if $P_i \in I$, send $x_i + r_i$ to all parties in $Q_i$, and run \func{VSS-Share} to secret-share $r_i$ in $Q_i$.
				
				\item If $P_i \notin I$, 
					\begin{enumerate}
						\item Choose $r_i$ and $x_i$ uniformly at random from $\F$ and \mbox{$\mask{x}_i \gets x_i + r_i$}.
						\item Send $\mask{x}_i$ to all parties in $Q_i$.
						\item Run \func{VSS-Share} to secret-share $r_i$ in $Q_i$. 
						\item For every party in $I_i$, add his share of $r_i$ and $\mask{x}_i$ to his view.
					\end{enumerate}
			
				\item For every party in $Q_i$, run \alg{Wait-For-Inputs} to wait for at least $n-t$ inputs.
				\begin{enumerate}
					\item Run \func{Thresh-Count} with flag $b_i$ initially set to zero to count the number of received inputs.
					
					\item If $x_i$ and $r_i$ are valid and consistent (based on the broadcast protocol and the verification stage of \alg{VSS-Reconst} respectively), raise an event to set $b_i \gets 1$ in \func{Thresh-Count}.
					
					\item Upon receiving \done from the parent quorum, run \alg{CMPC} using $b_i$ as your input to set $d \gets \textsf{True}$ if a $5/8$-fraction of the parties in $Q$ have their $b_i$'s set to one. Otherwise, $d \gets \textsf{False}$.
					
					\item If $d$ is set to $\textsf{False}$, then set $y \gets \textsf{Default}$ and $r_i \gets 0$.
				\end{enumerate}
		\end{enumerate}
	\end{algorithm}

	Let $V_1$ denote the view of the adversary from the hybrid execution, and $V_2$ be its view from the simulation. The inputs to \textsf{Thresh-Count} and Line~\ref{ln:waitThreshCMPC} of Protocol~\ref{pro:waitForInputs} are completely independent of the inputs of Protocol~\ref{pro:main}. Thus, $V_1$ contains only the masked inputs, $\mask{x_i}$'s, and at most $1/8$ fraction of the shares for each random mask, $r_i$'s. The masked inputs convey no information about the inputs. Moreover, a $1/8$ fraction of the shares are not enough to reconstruct the random number. Since $V_2$ contains all random elements, the adversary cannot distinguish $V_1$ from $V_2$.
	Since our simulator is straight-line and black-box, it follows from Theorem~\ref{thm:ucperfect} that the Input Commitment stage is UC-secure. \qed
\end{proof}

\subsection{Security of Circuit Evaluation}
We first prove the security of \alg{Gen-Rand}. The ideal functionality \func{Gen-Rand} is given in Protocol~\ref{fun:GenRand}. At least $7n/8$ of the inputs $\rho_1,...,\rho_N$ are sent by honest parties and thus are chosen uniformly and independently at random from $\mathbb{F}$. Hence, $r=\sum_{i=1}^{N}\rho_i$ is also a uniform and independent random element of $\mathbb{F}$. This is because the sum of elements of $\F$ is uniformly random if at least one of them is uniformly random.

\begin{algorithm}
	\algfont
	\caption{\func{Gen-Rand}}
	\label{fun:GenRand}
	
	\textit{Goal.} For a gate $u \in C$, generate a random value $r \in \mathbb{F}$ and VSS-share it among parties $P_1,...,P_N$ in the quorum associated with $u$. \medskip
	
	\textit{Functionality:}
	\begin{enumerate}			
		\item Receive inputs $\rho_1,...,\rho_N\in \mathbb{F}$ from $P_1,...,P_N$ respectively. For every $i\in[N]$, if $P_i$ does not send an input, then define $\rho_i = 0$.
		
		\item Calculate $r=\sum_{i=1}^{N}\rho_i$ and invoke \func{VSS-Share} to send a share $r_i$ of $r$ to $P_i$.
	\end{enumerate}
\end{algorithm}

\begin{lemma}
	The protocol \alg{Gen-Rand} is UC-secure.
\end{lemma}
\begin{proof}	
	We prove the $t$-security of \alg{Gen-Rand} in the \func{VSS-Share}-hybrid model which is similar to Protocol~\ref{pro:GenRand} except that every call to \alg{VSS-Share} is replaced with a call to the ideal functionality \func{VSS-Share}. The corresponding simulator \simu{Gen-Rand} is given in Protocol~\ref{pro:GenRandSim}.	
	\begin{algorithm}
		\algfont
		\caption{\simu{Gen-Rand}}
		\label{pro:GenRandSim}
		
		\textit{Inputs.} For a gate $u \in C$, the inputs $\{\rho_j\}_{P_j \in I_u}$ of the corrupted parties $P_1,...,P_N$ in the quorum associated with $u$.
		\medskip
		
		\textit{Simulation:}
		
		\begin{enumerate}
			\item For every \mbox{$P_i \in (Q_u - I_u)$} (\ie, for every honest party $P_i$), call \func{VSS-Share} with dummy input 0. Let $s^i_1,...,s^i_N$ denote the outputs.
			\item For every $P_j \in I_u$, 
			\begin{enumerate}
				\item Run \func{VSS-Share} with input $\rho_j$. Let $\rho^j_1,...,\rho^j_N$ denote the outputs. For every $k \in [N]$, add $\rho^k_j$ to the view of $P_j$.
				\item Compute $r_j=\sum_{k=1}^{N}\rho^k_j$ and add $r_j$ to the view of $P_j$.
			\end{enumerate}
		\end{enumerate}
	\end{algorithm}
	
	The views of the corrupted parties in the hybrid execution and the simulation are indistinguishable because the only difference between the two views is that \simu{Gen-Rand} generates the shares from dummy input 0 instead of actual inputs. Since \func{VSS-Share} generates uniform and independent random shares from any input, the two views are identically distributed.	
	Since our simulator is straight-line and black-box, \alg{Gen-Rand} is UC-secure.
	\qed
\end{proof}

We now proceed to the security proof of \alg{Circuit-Eval}. The ideal functionality \func{Circuit-Eval} is given in Protocol~\ref{fun:CircuitEval}.

\begin{algorithm}
	\algfont
	\caption{\func{Circuit-Eval}}
	\label{fun:CircuitEval}
	
	\textit{Goal.} For each gate $u \in C$ with children $v,w \in C$, $3N$ parties $P_1,...,P_{3N}$ provide inputs to the functionality to allow it evaluate the functionality of $u$ denoted by $f_u$. \medskip
	
	\textit{Functionality:}
	\begin{enumerate}			
		\item For every $i \in [N]$, receive $\rho_i$ from $P_i$, $\mask{y}_v$ and $r^{(i)}_v$ from $P_{i+N}$, and $\mask{y}_w$ and $r^{(i)}_w$ from $P_{i+2N}$ respectively.
		\item Run \func{Gen-Rand} with inputs $\rho_1,...,\rho_N$ to generate $r^{(1)}_u,...,r^{(N)}_u$. 
		\item Run \func{CMPC} to locally compute the following functionality: \label{ln:cmpcfunc}
		\begin{enumerate}
			\item $r_u \gets$ \func{VSS-Reconst} over $r^{(1)}_u,...,r^{(N)}_u$.
			\item $r_v \gets$ \func{VSS-Reconst} over $r^{(1)}_v,...,r^{(N)}_v$.
			\item $r_w \gets$ \func{VSS-Reconst} over $r^{(1)}_w,...,r^{(N)}_w$. 
			\item $y_1 \gets \mask{y}_v - r_v$
			\item $y_2 \gets \mask{y}_w - r_w$
			\item $\mask{y}_u \gets f_u(y_1,y_2) + r_u$
		\end{enumerate}
		
	\end{enumerate}
\end{algorithm}

\begin{lemma} \label{lem:ceuc}
	The protocol \textsf{Circuit-Eval} is UC-secure.
\end{lemma}
\begin{proof}
	We first show that for each gate $u \in C$, \func{Circuit-Eval} correctly computes $\mask{y}_u = y_u + r_u$. Based on \func{Input} and \func{Gen-Rand}, for each gate $u \in C$, the inputs of the honest parties in $Q_u$ are enough to reconstruct $r_u$. If $u$ is an input gate not included in the computation from the Input Commitment stage, then $r_u$ and its shares are 0. Thus, all three values of  $r_u$, $r_v$, and $r_w$ can be correctly reconstructed by the functionality since \func{\vssRec} can tolerate up to a $1/4$ fraction of the inputs being invalid.
	 
	We prove $\mask{y}_u = y_u + r_u$ by induction on the height of $u$, where $y_u$ is the correct output of the gate $u$. The base case is correct because based on the correctness of \func{Input}, for each input gate $v'$, we have $\mask{y}_{v'} = y_{v'} + r_{v'}$ and $r_{v'}$ can correctly be reconstructed from the inputs received from honest parties in $Q_{v'}$. Suppose that for all gates $u'$ whose height is less than the height of $u$, the functionality can compute $\mask{y}_{u'} = y_{u'} + r_{u'}$ and $r_{u'}$. This induction hypothesis is valid for $v$ and $w$.
	
	We now describe the induction step. In the computation of $u$, the functionality runs \func{CMPC}. We now argue based on the definition of the function computed by \func{CMPC} that the output of \func{CMPC} is $\mask{y}_u = r_u + y_u$. By the induction hypothesis, the functionality can reconstruct correct $r_v$ and $r_w$ and consequently it can correctly find $y_v$ and $y_w$ even if a $\fbad$ fraction of the inputs are missing. It is because the majority of the parties in $Q_v$ and $Q_w$ hold correct values of $\mask{y}_v$ and $\mask{y}_w$. Thus, the functionality can correctly compute $f_u(y_v,y_w) + r_u$. 

	We now prove the $t$-security of \alg{Circuit-Eval} in the (\func{Gen-Rand},\func{CMPC})-hybrid model which is similar to Protocol~\ref{pro:ce} except that every call to \alg{CMPC} and \alg{Gen-Rand} is replaced with a call to \func{CMPC} and \func{Gen-Rand} respectively. The corresponding simulator \simu{Circuit-Eval} is given in Protocol~\ref{pro:ceSim}.
	
	\begin{algorithm}
		\algfont
		\caption{\simu{Circuit-Eval}}
		\label{pro:ceSim}
		
		For every gate $u \in C$ with children $v,w \in C$, consider three groups of parties $Q_u, Q_v$, and $Q_w$, each of whom have $N$ parties. In each group, up to $N/8$ parties are corrupted.\smallskip
		
		\textit{Inputs.} $\{\rho_i\}_{P_i \in I_u}, \{r_u^{(i)}\}_{P_i\in (I_v \cup I_w)}$, and $\mask{y}_v$ and $\mask{y}_w$ from parties in $I_v \cup I_w$. \medskip
		
		\textit{Simulation:}
		
		\begin{enumerate}
			\item Run \func{Gen-Rand} with the following inputs: $\rho_i$ for every $P_i \in I_u$ and a dummy input for every party in $Q_u - I_u$. Let $\{r_u^{(i)}\}_{P_i\in Q_u}$ denote the outputs. For every $P_i \in I_u$, add $r_u^{(i)}$ to the view of $P_i$.
			
			\item Let $Q_\triangle = Q_u \cup Q_v \cup Q_w$ and $I_\triangle = I_u \cup I_v \cup I_w$. Run \func{CMPC} to compute the functionality defined in Line~\ref{ln:cmpcfunc} of \func{Circuit-Eval} with the following inputs: the input of every party in $I_\triangle$ as described in \func{Circuit-Eval}, and a dummy input for every party in $Q_\triangle - I_\triangle$. Let $\mask{y}_u$ denote the output. For every party in $I_\triangle$, add $\mask{y}_u$ to the view of the party.
		\end{enumerate}
	\end{algorithm}
	
	We now show that the views of the corrupted parties in the hybrid execution and the simulation are indistinguishable.
	After the evaluation of $u$, the following information will be added to the view of every corrupted party $P_i \in I_\triangle$: $\mask{y}_u$ and $\{r^{(j)}_u\}_{P_j \in I_u}$. Recall that $\mask{y}_u$ is the output of \func{CMPC} during the computation of $u$ which is equal to $y_u + r_u$, and $r_u$ is a uniformly random element of $\F$ based on \func{Gen-Rand}, independent of all other randomness in the algorithm.  	
	
	First, if a corrupted party $P_i$ is not in any of the quorums associated with $u,v$, and $w$, then no additional information will be added to its view during the computation of $u$; thus, its view will be identically distributed in the hybrid execution and the simulation.  
	
	Second, a corrupted party $P_i \in I_\triangle$ may add a share $r_u$ as well as shares of the individual random elements whose sum is $r_u$ to its view in the computation of \func{Gen-Rand}. Also, it adds $y_u+r_u$ to its view. However, $P_i$ cannot learn any additional information about the shares of $r_u$ (and thus about $r_u$) based on \func{CMPC} and \func{Gen-Rand}. In other words, the parties in $I_\triangle$ are unable to directly determine $r_u$, since the only relevant inputs are the shares of $r_u$, and they do not have enough of those since they have fewer than half of them. 
	
	These parties also do not have enough shares of shares of $r_u$ to reconstruct it. However, they add to their view shares of each of the other shares of $r_u$ multiple times: once during the input stage of \func{CMPC} in which $u$ is involved, and once during the computation of the parent of $u$. Each time, they do not get enough shares of shares $r_u$ to reconstruct any shares of $r_u$. But, can they combine the shares of shares from different runs for the same secret to gain some information? Since fresh and independent randomness was used by the dealers creating these shares on each run, the shares from each run are independent of the other runs, and so they do not collectively give any more information than each of the runs give separately. Since each run does not give the parties in $I_\triangle$ enough shares to reconstruct anything, it follows that they do not learn any information about $r_u$.
	
	Second, parties in $I_\triangle$ add shares of shares for $r_v$ and $r_w$ to their views. However, with a similar argument as $r_u$, they cannot reconstruct $r_v$ and $r_w$ as well even if these parties participate in one or more of the instances of \func{CMPC} which involve $v$ or $w$: the computation of $v$ or $w$ themselves or the computations of $u$ as their parents.
	
	Moreover, $\mask{y}_u$ is also a random element in the field since $r_u$ is uniformly random and $\mask{y}_u= y_u + r_u$. Thus, $\mask{y}_u$ holds no information about $y_u$, and the corrupted parties cannot learn any information about $y_u$ except what is implicit in his input and the circuit output. This means that the corrupted parties cannot distinguish if they are participating in a run of the hybrid model or the simulation. Finally, since \simu{Circuit-Eval} is straight-line and black-box, \alg{Circuit-Eval} is UC-secure.
	\qed
\end{proof}

\subsection{Security of Output Stages} \label{sec:proof-output}
The ideal functionality for the Output Reconstruction and the Output Propagation stages of Protocol~\ref{pro:main} are given in Protocol~\ref{fun:Output}.

\begin{algorithm}
	\algfont
	\caption{\func{Output}}
	\label{fun:Output}
	
	\textit{Goal.} The functionality guarantees the output is reconstructed correctly and it is learned by all honest parties.
	\medskip
	
	\textit{Functionality:}
	\begin{enumerate}
		\item Run \func{VSS-Reconst} to reconstruct the output.
		\item Send the output to all the parties.		
	\end{enumerate}
\end{algorithm} 

\begin{lemma}
	The Output Reconstruction and Output Propagation stages of Protocol~\ref{pro:main} are UC-secure.
\label{lem:outputuc}	
\end{lemma}
\begin{proof}
	We first show that the two stages correctly compute \func{Output}. Let $z$ be the output gate of $C$. By Lemma~\ref{lem:ceuc}, all parties in the output quorum $Q_z$ eventually agree on $y_z + r_z$ and hold shares of $r_z$. In the Output Reconstruction stage, these parties run the \vssRec. Since at least a $\fgoodq$ fraction of them are honest, they correctly reconstruct $r_z$. Since all honest parties in $Q_z$ know $y_z + r_z$ and subtract from it the reconstructed $r_z$, they all eventually learn $y_z$. Thus, all parties in $Q_z$ eventually learn $y_z$.

	We now show by induction that all honest parties eventually learn $y_z$. Since $Q_1$ is assigned to the output gate, it provides a base case. For $i > 1$, consider the parties in $Q_i$, and for all $j < i$ assume the correct output is learned by all parties in $Q_j$. During the Output Propagation stage, the parties in $Q_i$ receive putative values for the output from the parties at $Q_{\lfloor i/2 \rfloor}$. Since $Q_{\lfloor i/2 \rfloor}$ is good, and by induction hypothesis all honest parties in it have learned the correct output, it follows that all honest parties in quorum $Q_{\lfloor i/2 \rfloor}$ send the same message which is the correct output. By Lemma~\ref{lem:qcomm}, all honest parties in $Q_i$ eventually learn the correct output. By induction, all the parties learn the correct value.
	
	We now prove the $t$-security of the output stages in the \func{VSS-Reconst}-hybrid model. The corresponding simulator \simu{Output} is given in Protocol~\ref{pro:output}.		
	\begin{algorithm}
		\algfont
		\caption{\simu{Output}}
		\label{pro:output}
		\smallskip
		\textit{Inputs.} For the output gate $z$ and the corresponding quorum $Q_z$, the inputs of the simulator are $\{r_z^{(i)}\}_{P_i\in I_z}$, and $\mask{y}_z$ from parties in $I_z$. \medskip
		
		\textit{Simulation:}
		
		\begin{enumerate}
			\item Run \func{VSS-Reconst} with inputs  $\{r_z^{(i)}\}_{P_i\in I_z}$ and dummy inputs for honest parties. Add the output to the view of parties in $I_z$.
		 
			\item For every $i \in \{2,...,n\}$, parties in $Q_i$ perform the following steps:
					\smallskip
					\begin{enumerate}
						\item Receive $y$ from $Q_{\lfloor i/2\rfloor}$ and add it to the view of every parties in $I_{\lfloor i/2\rfloor}$.
						\item Send $y$ to all parties in $Q_{2i}$ and	$Q_{2i+1}$.
					\end{enumerate}		
		\end{enumerate}
	\end{algorithm}		
	The views of the corrupted parties in the hybrid execution and the simulation are indistinguishable since the only message that is added to the view of the adversary is the output. Based on the security definition of MPC, the adversary is allowed to learn the output.
	\qed
\end{proof}

\subsection{Security of Protocol~\ref{pro:main}}
We now show that our main protocol is UC-secure.

\begin{lemma} \label{lem:step2}
	Protocol~\ref{pro:main} is UC-secure.
\end{lemma}
\begin{proof}
	Canetti~\cite{canetti:studies} proves the $t$-security of \alg{VSS-Share}, \alg{VSS-Reconst}, and \alg{CMPC} using straight-line black-box simulators. So, based on Theorem~\ref{thm:ucperfect}, these protocols are UC-secure. Moreover, Lindell~\etal~\cite{Lindell:2006:CAB:1217856.1217857} show that any Byzantine agreement protocol in the standard model (such as the protocol of~\cite{canetti}) is UC-secure. Hence, the Byzantine agreement of~\cite{Feldman:1988:OAB:62212.62225} is also UC-secure.
	
	Protocol~\ref{pro:main} is $t$-secure since in lemmas~\ref{lem:inputuc},~\ref{lem:ceuc}, and~\ref{lem:outputuc} we showed that all stages of the Protocol~\ref{pro:main} are $t$-secure.
	Based on Theorem~\ref{thm:ucperfect}, since we have proved the \mbox{$t$-security} of Protocol~\ref{pro:main} using a straight-line black-box simulator, the protocol is also UC-secure.
	\qed
\end{proof}

%
%
%
%

\subsection{Cost Analysis} \label{sec:costAnalysis}
We now analyze the resource costs of Protocol~\ref{pro:main}.

\begin{lemma} \label{lem:iccost} 
	During the Input stage, each quorum sends at most $O(\log n)$ messages.
\end{lemma}
\begin{proof}
	For the input stage, each quorum is mapped to at most one of the input gates and hence one of the nodes in the count tree. Thus, from Theorem~\ref{thm:taucount} it follows that the total number of messages sent by each quorum is $O(\log n)$. Since each quorum has $\log{n}$ parties, an additional $\polylog(n)$ messages are sent by each quorum during  \vssSh and \vssRec to check whether the input is correctly secret-shared.
	\qed
\end{proof}

\begin{lemma}
	If all honest parties follow Protocol~\ref{pro:main}, then with high probability, each party sends at most $\tilde{O}(m/n+\sqrt{n})$ messages.
\end{lemma}
\begin{proof}
	By Theorem~\ref{thm:quorum-building}, we need to send $\tilde{O}(\sqrt{n})$ messages per party to build the quorums. Subsequently, each party must send messages for each quorum in which it is a member. Recall that each party is in $\Theta(\log n)$ quorums.
	
	By Lemma~\ref{lem:iccost}, each quorum sends $\tilde{O}(\log(n))$ messages during Input stage. Recall that each quorum is mapped to $\Theta\big(\frac{m+n}{n}\big)$ nodes of $C$.
	A quorum runs \textsf{Gen-Rand} and the gate evaluation step of \ce once per node it is mapped to in $C$.
	Since each gate has in-degree two and out-degree at most two, a quorum runs \hw at most three times for every node it is mapped to in $C$. Also, at most $\polylog(n)$ messages are sent per party per instance of \hw, \textsf{Gen-Rand}, and gate evaluation. Finally, each quorum sends $O(\log n)$ messages in the dissemination of the output. Thus, each quorum sends $\polylog(n)$ messages per node it represents. It follows that each party sends $\tilde{O}(m/n +\sqrt{n})$ messages.
	\qed
\end{proof}

\begin{lemma}
	If all honest parties follow Protocol~\ref{pro:main}, with high probability, the total latency is $O(d\polylog(n))$ where $d$ is depth of the circuit the protocol computes.
\end{lemma}
\begin{proof}
	Based on Theorem~\ref{thm:quorum-building}, the latency for creating quorums is $\polylog(n)$. Based on Theorem~\ref{thm:taucount}, the latency for the \alg{Thresh-Count} algorithm is $O(\log{n})$ which implies that the Input Commitment stage also has $\polylog(n)$ latency.
	
	In the computation of the circuit, to evaluate the gate $g$ in the upper level of the circuit, first its input gates in lower level of the circuit must be evaluated. This implies that the evaluation of the circuit is level by level and the latency for evaluating the circuit is $O(d)$ times the latency of \hw over $\log{n}$ parties.
	\qed
\end{proof}

\section{Asynchronous Threshold Counting}\label{sec:taucount}
In this section, we present an asynchronous Monte Carlo algorithm called \alg{Thresh-Count} which solves the threshold counting problem and provides the following theorem proved in Section~\ref{sec:tc-proofs}.

Our threshold counting algorithm runs in a setting with $n$ honest parties in a fully-connected network with private and authenticated channels and asynchronous communication. In our asynchronous MPC protocol presented in Section~\ref{sec:alg}, we run \alg{Thresh-Count} among a set of quorums, where each quorum represents an honest party.

\begin{theorem}
	The algorithm \alg{Thresh-Count} solves the threshold counting problem with high probability, while ensuring:
	\begin{enumerate}
		\item Each party sends at most $O(\log{n})$ messages of constant size,
		\item Each party receives at most  $O(\log{n})$ messages,
		\item Each party performs $O(\log{n})$ computations,
		\item Total latency is $O(\log{n})$.
	\end{enumerate}
	\label{thm:taucount}
\end{theorem}

Recall that in the threshold counting problem there are $n$ honest parties in an asynchronous communication network with private channels. Each party has an input flag which is initially $\mathsf{0}$. At least $\tau$ of the parties' bits will eventually be set to $\mathsf{1}$ based on an external event. When this happens, we say the threshold is reached. The goal is for each of the parties to terminate at some time after the threshold is reached.

Although in our application $\tau$ is linear in $n$, we address the more general case, where $\tau = O(n)$. Our algorithm depends on prior knowledge of $\tau$. As specified in Theorem~\ref{thm:taucount}, each party running the algorithm sends and receives $O(\log{n})$ messages of constant size and performs $O(\log n)$ computations; moreover the total latency is $O(\log{n})$.


For ease of presentation, we first describe an algorithm which works when $\tau = \Theta(n)$, in particular, when $\tau$ is at least $n/2$. We then indicate why this fails when $\tau$ is smaller, and show how to modify it so that it works for all $\tau$. The formal algorithm is shown as Protocol~\ref{pro:taucount}.

Consider a complete binary tree where each party sends its input to a unique leaf node when it is set to $1$. Then, for every node $v$, each child of $v$ sends $v$ a message giving the number of inputs it has received so far and it sends a new message every time this number changes. The problem with this approach is that it is not load-balanced: each node at depth $i$ has $n/2^i$ descendants in the tree, and therefore, in the worst case, sends and receives $n/2^i$ messages. Thus, a child of the root sends $n/2$ messages to the root and receives the same number of messages from its children. 

To solve the load-balancing problem, we use a randomized approach which ensures with high probability that each leaf of the data structure receives at least $7\log n$ messages and does not communicate with its parent until it has done so. Subsequent messages it receives are not forwarded to its parent but rather to other randomly chosen leaves to ensure a close to uniform distribution of the messages.	

\begin{figure*}[t]
	\begin{center}
		\includegraphics[scale=0.19]{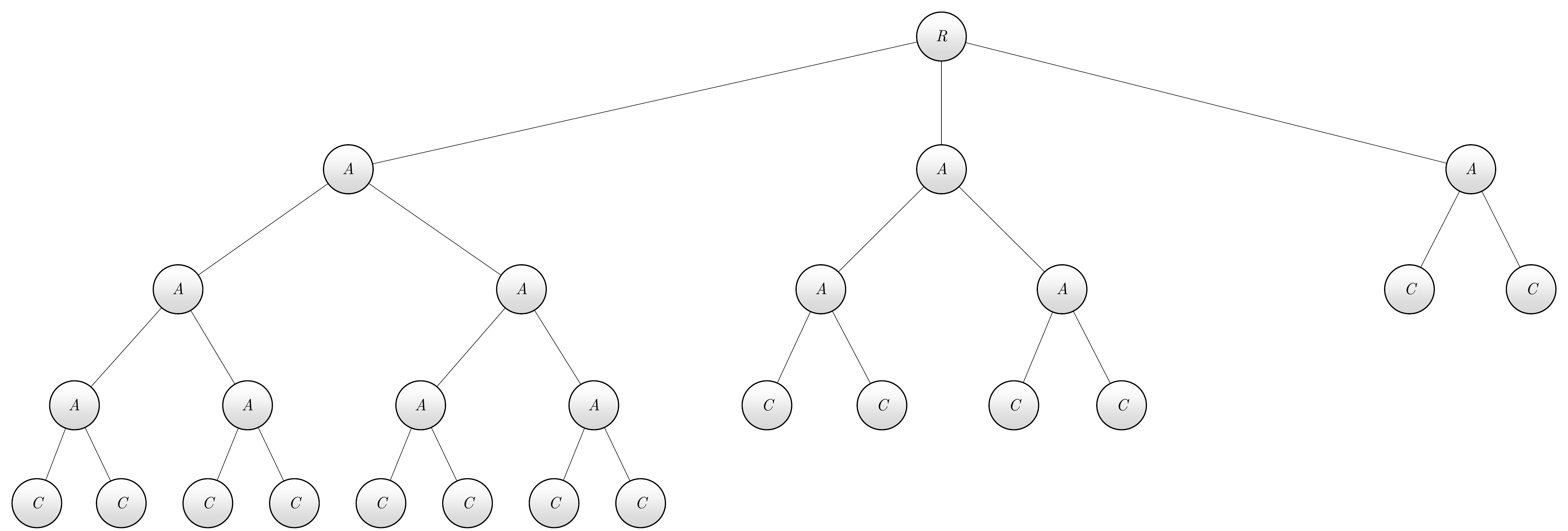}
	\end{center}
	\caption{The \ct for $n=2048$ and $\tau=1232$. $\dcl = \lceil \log{\frac{1232}{14 \times 11}}\rceil = 3$. The node marked $R$ is the root, nodes marked $A$ are adding nodes, and nodes marked $C$ are collection nodes.}
	\label{fig:tc}
\end{figure*}

Our algorithm consists of up and down stages. For the up stage the parties are arranged in a predetermined tree data structure, which we call the \emph{\ct.} The \ct consists of a root node with $O(\log{n})$ children, each of which is itself the root of a complete binary tree; these subtrees have varying depths as depicted in Figure~\ref{fig:tc}. In the up stage, parties in the trees count the number of \oneinputs, \ie, the number of parties' inputs that are set to $\mathbf{1}$. The root then eventually decides when the threshold is reached. In the down stage, the root notifies all the parties of this event via a complete binary tree of depth $\log n$. Note that the tree used in the down stage has the same root as the \ct.

Let $\dcl = \lceil \log{\frac{\tau}{14\log{n}}}\rceil$. Note that $\dcl = O(\log{n})$. The root of the \ct has degree $\dcl$. Each of the $\dcl$ children of the root is itself the root of a complete binary subtree, which we will call a \emph{collection subtree}. For $1\le j\le \dcl$, the $j$th collection subtree has depth $\dcl+1-j$. Party 1 is assigned to the root and parties 2 to $\dcl+1$, are assigned to its children, \ie, the roots of the collection subtrees, with party $j+1$ being assigned to the $j$th child. The remaining nodes of the collection trees are assigned parties in order, starting with $\dcl+2$, left to right and top to bottom. One can easily see that the entire data structure has fewer than $n$ nodes, (in fact it has fewer than $\frac{\tau}{3\log{n}}$ nodes) so some parties will not be assigned to any node.

The leaves of each collection subtree are \emph{collection nodes}, while the internal nodes of each collection tree are \emph{adding nodes}. 

\subsection{Up Stage}
When a party's input is set to $\mathbf{1}$, it sends a \flag message, which we will sometimes simply refer to as a flag, to a uniformly random collection node from the first collection subtree. Intuitively, we want the flags to be distributed as evenly as possible among the collection nodes. The parameters of the algorithm are set up so that with high probability each collection node receives at least $7\log{n}$ \flag messages.

Each collection node in the $j$-th collection tree waits until it has received $7\log{n}$ flags.
It then sends its parent a \countm message. For each additional flag received,  up to $14 \log n$, it chooses a uniformly random collection node in the $(j+1)$-st collection subtree and forwards a flag to it. If $j = \dcl$, then it forwards these $14\log{n}$ flags directly to the root.
Subsequent flags are ignored. Again, we use the randomness to ensure a close to even distribution of flags with high probability.

Each adding node waits until it has received a \countm message from each of its children. Then, it sends a \countm message to its parent. We note that, with high probability, each adding node sends exactly one message during the algorithm.
The parameters of the algorithm are arranged so that all the \countm messages that are sent in the the $j$th collection subtree together account for $\tau/2^j$ of the \oneinputs. Thus, all the \countm messages in all the collection subtrees together account for $\tau \left(1-\frac1{2^{\dcl}}\right)$ of the \oneinputs. At least $\frac{\tau}{2^{\dcl}}$ \oneinputs remain unaccounted for. These \oneinputs and up to $O(\log{n})$ more are collected as flags at the root.

\subsection{Down Stage}

When party 1, at the root,  has accounted for at least $\tau$
\oneinputs, it starts the down stage by sending the \done message to parties 2 and 3. For $j>1$, when party $j$ receives the \done message, it forwards this message to parties $2j$ and $2j +1$. Thus, eventually the
\done message reaches all the parties, who then learn that the threshold has been met.

Note that all three types of messages sent in this protocol, \flag, \countm and \done, are notifications only; they do not contain any numerical value.
Since 2 bits are sufficient to distinguish three different kinds of messages, all the messages sent in this protocol are 2-bit strings. Note that we distinguish between flags and \countm messages since the root receives both kinds. However it is the only node for which this is a problem. We could add another node, as the $(D+1)$st child of the root, (equivalently as a collection subtree of depth 0,) which waits for $14\log{n}$ messages, and sends a \countm message to the root. In so doing, we could eliminate the need to explicitly distinguish \flag and \countm message, since they would be automatically distinguished by the role of the receiving node. Thus, we could actually reduce all message lengths to a single bit.

\subsection{Handling Sublinear Thresholds}
Now, we consider the case where $\tau = o(n)$. It is easy to see that the worst load in terms of the number of received messages is when all $n$ inputs are $\mathsf{1}$.
In this case, a collection node in the first collection subtree receives, on average, $14( n/\tau)\log{n}$ flags. When $\tau = \Theta(n)$, this is still $O(\log{n})$, but when $\tau=o(n)$ this is $\omega(\log{n})$. Before we describe how to fix this, we note that the problem exists \emph{only} in the leaves of the \emph{first} collection subtree. Subsequent collection nodes receive only $O(\log{n})$ flags, because each node only forwards up to $14\log{n}$ flags.

\begin{figure*}[t]
	\begin{center}
		\includegraphics[scale=0.19]{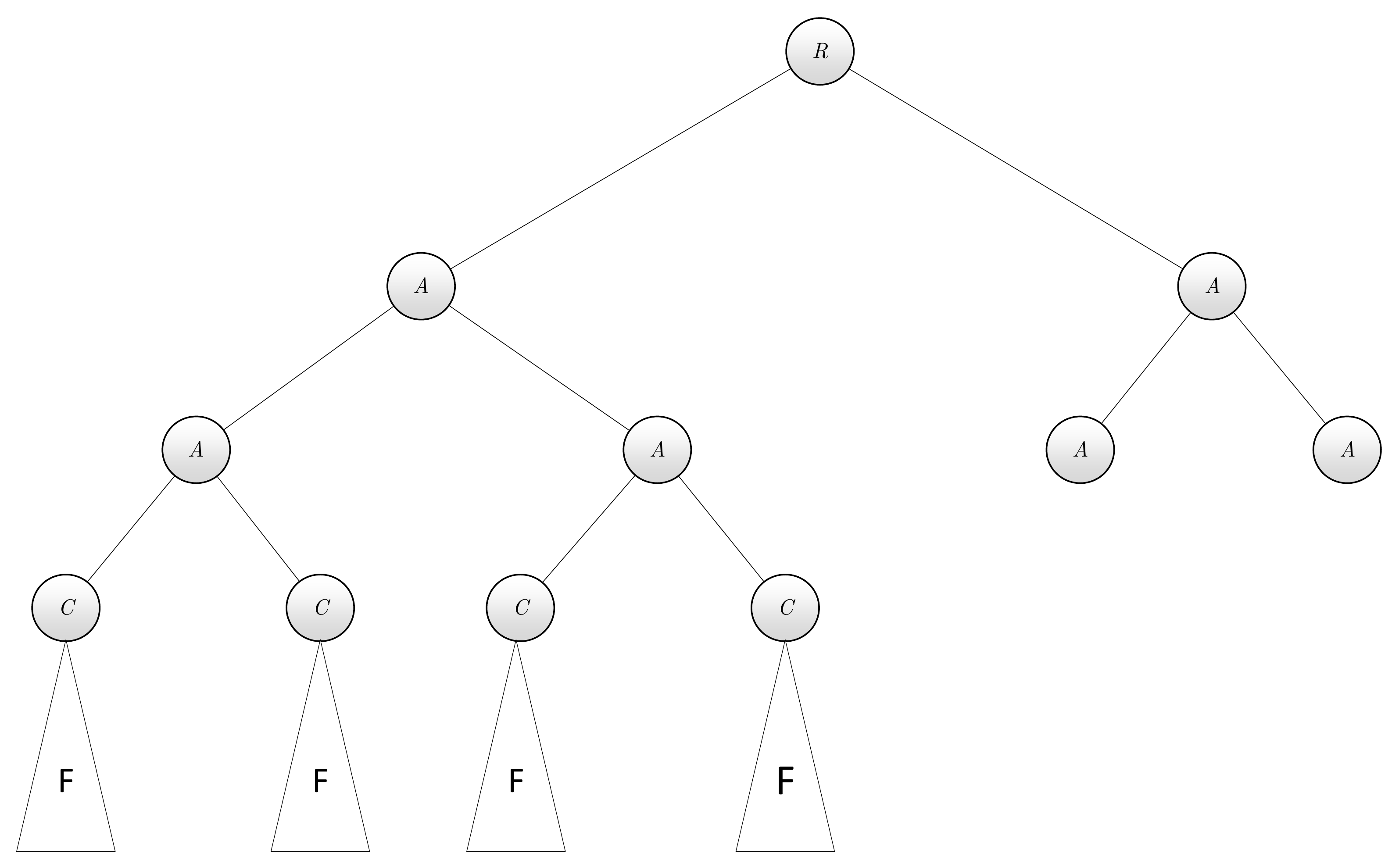}
	\end{center}
	\caption{The \ct for $n=2048$ and $\tau=616$.
		$\dcl = \lceil \log{\frac{616}{14 \times 11}}\rceil = 2$.
		The node marked $R$ is the root, nodes marked $A$ are adding
		nodes and nodes marked $C$ are collection nodes. The filters,
		marked $F$, are complete binary trees of depth 7, with 128 leaves each, for a total of 512 filter leaves.
		}\label{fig:tcfilter}
\end{figure*}

For the sake of having a definite cutoff and tractable constants, we will apply the following fix whenever $\tau < n/2$. Below each collection node in the first collection tree, we put in a \emph{filter}, which is a complete binary tree of depth $\log{n} -2 -D$ with $\frac{7n\log{n}}{2\tau}$ leaves. This is equivalent to extending the first collection tree to depth $\log{n}-2$ so that it has $n/4$ leaves. The collection nodes will remain at depth $D$ though. See Figure~\ref{fig:tcfilter}.

When a party's input is set to $\mathsf{1}$, it selects a random collection node in the first collection tree, but rather than sending a flag directly to it, it sends the flag to a random leaf of the collection node's filter. The nodes in the filter simply forward any flags they receive, up to $21\log{n}$, to their parent in the filter. Subsequent flags are ignored. Clearly, this means that the collection node at the root of the filter cannot receive more than $42 \log{n}$ flags, which solves the load problem. Moreover, we have not simply transfered the problem to the leaves of the filter. Since there are so many more of them, each one actually receives fewer flags on average and the parameters are adjusted to make their maximum load $O(\log{n})$ with high probability. As we will also see in the analysis, these filters do not filter out too many flags; when there are only $\tau$ \oneinputs among the parties, with high probability all the flags get through.

\begin{algorithm}
	\caption{\alg{Thresh-Count}}
	\medskip
	\algfont
	\textit{Goal.} $n$ is the number of parties, $\tau$ is the threshold, $b$ is a flag bit initially set to zero which may be set to one by an external event throughout the protocol and $\dcl = \lceil \log(\frac{\tau}{14\log{n}})\rceil$. The algorithm notifies all the parties upon receiving $\tau$ flag bits set to one.

	\begin{enumerate}
		\item \textbf{Setup.} No messages sent in this stage:
		\begin{enumerate}
			\item Build the count tree and set party 1 as the root:\\
			 For $1\le j \le \dcl$, party $j+1$ is a child of the root (and the root of the $j$th collection subtree with depth $\dcl+1-j$). Starting with party $\dcl+2$, the remainder of the nodes are assigned to parties, left to right and top to bottom. If $\tau < n/2$ the remaining parties are assigned to filters, left to right and top to bottom.
			\item Let $sum = 0$ for the root.
		\end{enumerate}
		\item \textbf{Up Stage.}
		\begin{enumerate}
			\item Upon $b=1$, choose a uniformly random collection node $v$ from collection subtree 1,
			\begin{itemize}
				\item If $\tau > n/2$, send a \flag to $v$.
				\item Otherwise, choose a uniformly random leaf in $v$'s filter and send a \flag to it.
			\end{itemize}
			\item Upon receiving a \flag, if previously forwarded
			fewer than $21\log{n}$ flags, forward the flag to parent. Otherwise, ignore it.
			\item Perform the following steps to collect nodes in the collection subtree $j$:
			\begin{itemize}
				\item Upon receiving $7\log{n}$ {\flag}s,
				send parent a \countm message.
				\item Upon subsequently receiving a \flag, if $j<\dcl$, send it to a uniformly random collection node in collection subtree $j+1$.
				If $j=\dcl$, then send it directly to the root. Do this for up to $14 \log n$ flags. Then, ignore all subsequent \flag messages.
			\end{itemize}
			\item Upon receiving \countm from both children, send \countm to the parent.
			
			\item If $sum < \tau$,
			\begin{itemize}
				\item Upon receiving a \countm from party $j+1$, set $sum \gets sum + \tau/2^j$.
				\item Upon receiving a \flag, $sum \gets sum+1$.
			\end{itemize}
		\end{enumerate}
		\item \textbf{Down Stage.} If $sum \ge \tau$,
		\begin{enumerate}
			\item Party 1 (the root): Send \done to parties 2 and 3, and then terminate.
			\item Party $j$ for $j>1$: Upon receiving \done from party $\lfloor j/2 \rfloor$, forward it to parties $2j$ and $2j+1$ (if they exist), and then terminate.
		\end{enumerate}
	\end{enumerate}
	\label{pro:taucount}
\end{algorithm}

\subsection{Proof of Theorem~\ref{thm:taucount}} \label{sec:tc-proofs}
In this section, we prove the correctness and resource costs of Protocol~\ref{pro:taucount}.
The process of each party independently selecting a random collection node to notify after its input has been set to $\mathbf{1}$ can be modeled as a balls and bins problem and hence be approximated by the Poisson distribution.

\subsubsection{Preliminaries}
We first recall the following Chernoff bound for a Poisson random variable from Mitzenmacher and Upfal~\cite{opac-b1117540}. 

\begin{theorem}[Theorem 5.4 of \cite{opac-b1117540}]\label{thm:CherPois}
	Let $Y \sim$ Poisson$(\mu)$. Then,
	\begin{enumerate}
		\item for $x < \mu$, $\Prob(Y \le x)\le \e^{-\mu}(\e \mu/x)^{x}$, and
		\item for $x > \mu$, $\Prob(Y \ge x)\le \e^{-\mu}(\e \mu/x)^{x}$.
	\end{enumerate}
\end{theorem}

\begin{lemma} \label{lem:ballsnbins}
	Assume $\alpha \bins$ balls are thrown independently and uniformly at random into $\bins$ bins. Let $E_1$ denote the event that the minimum load is less than $\alpha/2$, and let $E_2$ denote the event that the maximum load exceeds $3\alpha/2$. Then,
	\begin{equation}\label{eqn:bb-low}
	\Prob (E_1) \le \e\bins\sqrt{\alpha \bins}
	\left(\frac2{\e}\right)^{\alpha/2}
	\end{equation}
	and
	\begin{equation}\label{eqn:bb-high}
	\Prob (E_2) \le \e\bins\sqrt{\alpha \bins}
	\left(\frac{8\e}{27}\right)^{\alpha/2}\,.
	\end{equation}
\end{lemma}
\begin{proof}
	For $1\le i \le \bins$, let $X_i$ denote the number of balls in the $i$th bin, and let $Y_i \sim$ Poisson$(\alpha)$ be an independent Poisson random variable with mean $\alpha$. It is well known that the distribution of each $X_i$ is close to that of $Y_i$, and moreover that the
	\emph{joint} distribution of the $X_i$'s is well approximated by the joint (\ie, product) distribution of the $Y_i$'s (see Chapter 5 in Mitzenmacher and Upfal~\cite{opac-b1117540}). Indeed, Corollary 5.11 from~\cite{opac-b1117540} states that for any event $E$ that is monotone in the number of balls, if $E$ occurs with probability at most $p$ in the Poisson approximation, then $E$ occurs with probability at most $2p$ in the exact case. Since maximum and minimum load are both clearly monotone increasing in the number of balls, applying this corollary we have:
	\begin{align*}
		\Prob(E_1)&= \Prob\left(\exists i \mbox{ s.t. } X_i \le \alpha/2\right) \\
		& \le 2\Prob\left(\exists i \mbox{ s.t. } Y_i \le
		\alpha/2\right)\\
		&\le 2 \sum_{i=1}^{\bins}
		\Prob\left(Y_i \le \frac{\alpha}{2}\right) \\
		&\le 2 \bins\left(\frac{2}{\e}\right)^{\alpha/2},
	\end{align*}
	where the last inequality follows from Theorem~\ref{thm:CherPois} with $\mu = \alpha$ and $x = \alpha/2$.
	Similarly,
	\begin{align*}
		\Prob(E_2)&= \Prob\left(\exists i \mbox{ s.t. } X_i > 3\alpha/2\right) \\
		&\le 2 \Prob\left(\exists i \mbox{ s.t. } Y_i >
		3\alpha/2\right)\\
		&\le 2 \sum_{i=1}^{\bins}
		\Prob\left(Y_i \ge \frac{3\alpha}{2}\right) \\
		&\le 2\bins \left(\frac{8\e}{27}\right)^{\alpha/2},
	\end{align*}
	where the last inequality follows from Theorem~\ref{thm:CherPois} with $\mu = \alpha$ and $x = 3\alpha/2$.
	\qed
\end{proof}

\subsubsection{Protocol Analysis}
Let $\sigma$ be the number of \oneinputs. We know that $\tau \le \sigma \le n$. Let $s=\sigma/\tau$.
For simplicity of the analysis, we will assume that the first $\tau$ flags to be sent are \emph{marked} while the remaining $\sigma -\tau$ are
\emph{unmarked}. As we track the progress of the flags through our data structure, we pay particular attention to the marked flags. Due to asynchrony, the marked flags need not be the first $\tau$ to arrive at their destinations.
\begin{lemma}\label{lem:bigtau}
	Suppose $\tau \ge n/2$.
	In the \alg{Thresh-Count} algorithm, with probability at least $1- \frac{1}{7n\log{n}}$, the first collection subtree satisfies all of the following:
	\begin{enumerate}
		\item Each collection node receives between $7s\log{n}$  and
		$21s\log{n}$ flags.
		\item The \countm messages generated in this tree, when they reach
		the root, account for $\tau/2$ \oneinputs.
		\item At least $\tau/2$ and at most $\tau$ flags are forwarded to the
		second collection tree.
	\end{enumerate}
\end{lemma}
\begin{proof}
	The process of sending $\sigma$ \flag messages to the collection nodes in the first collection tree can be modeled as a balls and bins problem as in Lemma~\ref{lem:ballsnbins}  with $\alpha = 14s\log{n}$ and $\bins =\tau /14\log{n}$.
	$E_1$ and $E_2$ are, respectively, the events that some collection node fails to receive $7s\log{n}$ flags and that some collection node receives more than $21s\log{n}$ flags.
	By applying the lemma, we get
	\vspace{-0.07em}
	\begin{align*}
		\Prob(E_1) &\le \frac{2\tau}{14\log{n}}\left(\frac2{\e}\right)^{7s\log{n}} \\
		&\le \frac{2 n}{14\log{n}} 2^{-0.4426 \times 7s\log n} \\
		&\le \frac{1}{7n^{2s}\log{n}}
	\end{align*}
	and
	\begin{align*}
		\Prob (E_2) &\le \frac{2 \tau}{14\log{n}}
		\left(\frac{8\e}{27}\right)^{7s\log{n}} \\
		&\le \frac{2 n}{14\log{n}}2^{-0.3121 \times 7s\log{n}} \\
		&\le \frac{1}{7n^{1.1s}\log{n}}
	\end{align*}
	Thus, the probability that (a) fails is at most $\frac{1+n^{0.9s}}{7n^{2s}\log{n}}$.
	
	To see (b), we note that there are $\tau/(14\log{n})$ collection nodes in the first collection subtree, each of whom generates a \countm message when it has received $7\log{n}$ flags. The flags correspond to distinct \oneinputs, and hence together they account for $\tau/2$ \oneinputs.
	Thus, (b) fails only if some node fails to receive at least $7\log{n}$ flags, which is already accounted for in the failure of (a).
	
	To prove (c), we need to track the progress of the marked flags.
	Let $E'_1$ and $E'_2$ denote respectively, the events that some node fails to receive at least $7\log{n}$ marked flags and that some node receives more than $21\log{n}$ marked flags. Then, since there are $\tau$ marked flags, applying Lemma~\ref{lem:ballsnbins} with $\alpha = 14\log{n}$ and $\bins = \tau/14\log{n}$ we see that
	\begin{align*}
		\Prob(E'_1) &\le \frac{2\tau}{14\log{n}}\left(\frac2{\e}\right)^{7\log{n}} \\
		&\le \frac{2 n}{14\log{n}} 2^{-0.4426 \times 7\log n} \\
		&\le \frac{1}{7n^{2}\log{n}}
	\end{align*}
	and
	\begin{align*}
		\Prob (E'_2) &\le \frac{2 \tau}{14\log{n}} \left(\frac{8\e}{27}\right)^{7\log{n}}\\
		&\le \frac{2 n}{14\log{n}}2^{-0.3121 \times 7\log{n}} \\
		&\le \frac{1}{7n^{1.1}\log{n}}.
	\end{align*}
	
	Within each collection node, by transferring the marks from some marked flags to some unmarked flags, we may assume that the marked flags are the first to arrive. We can do this transfer because it does not change the distribution of marked and unmarked flags between the nodes, nor does it change the total number of marked flags across all collection nodes. The advantage of this change is that in following the algorithm, each node will first use all its marked flags before using unmarked flags.
	
	In particular, as long as $E'_1$ and $E'_2$ do not occur, each node will use $7\log{n}$ flags to generate a \countm message, after which it will be left with between 0 and $14\log{n}$ marked flags. Since it forwards up to $14\log{n}$ flags to the next collection subtree, it follows that it will forward \emph{all} of its marked flags and possibly some unmarked flags to the next subtree. Since there are $\tau$ marked flags across all the collection nodes, and the \countm messages account for $\tau/2$ of them, it follows that the remaining $\tau/2$ marked flags are forwarded. Hence, at least $\tau/2$ flags are forwarded. Moreover, since there are $\tau/(14\log{n})$ nodes and each forwards at up to $14\log{n}$ flags, at most $\tau$ flags are forwarded, which establishes (c).
	
	Now, let $E = E_1 \cup E_2 \cup E'_1 \cup E'_2$ be the union of all the bad events we've encountered.
	For large enough $n$,
	\begin{alignat*}{2}
	\Prob(E) &\le \frac{1}{7n^{2s}\log{n}} + \frac{1}{7n^{1.1s}\log{n}} + \frac{1}{7n^{2}\log{n}} \\
	&\quad\quad\quad + \frac{1}{7n^{1.1}\log{n}} \\
	&\le \frac{1}{7n\log{n}}
	\end{alignat*}
	Thus, with probability at least $1-\frac{1}{7n\log{n}}$, (a), (b), and (c) are all true, as desired.
	\qed
\end{proof}

We will also need to prove a similar lemma when $\tau < n/2$.
Note that when $\tau \ge n/2$, we have $\sigma \le 2\tau$, or $s = \sigma/\tau \le 2$. When $\tau < n/2$, $\sigma$ may be much bigger than $\tau$. Let $M = \min\{ \sigma/\tau, 2\}$.

\begin{lemma}\label{lem:smalltau}
	Suppose $\tau < n/2$.
	In the \alg{Thresh-Count} algorithm, with probability at least $1- \frac{1}{7n\log{n}}$, the first collection subtree satisfies all of the following:
	\begin{enumerate}
		\item Each collection node receives between $7\log{n}$  and
		$21M\log{n}$ flags.
		\item Each filter node receives at most $21M\log{n}$ flags.
		\item The \countm messages generated in this tree, when they reach
		the root, account for $\tau/2$ \oneinputs.
		\item At least $\tau/2$ and at most $\tau$ flags are forwarded to the
		second collection tree.
	\end{enumerate}
\end{lemma}
\begin{proof}
	When $\tau < n/2$, the flags are not sent directly to the collection nodes, but rather to leaf nodes of the filters below the collection nodes.
	We will say that a filter receives a flag if the flag is received by any of its leaf nodes.
	
	We first note that each party's process of selecting a random collection node, and then selecting a random leaf in its filter, is equivalent to simply selecting a uniformly random leaf node from among all the leaf nodes for all the filters. We've already remarked that adding the filters is equivalent to extending the first collection subtree to depth $\log{n} -2$ while keeping the collection layer the same. Thus, there are $n/4$ filter leaf nodes to choose from. Using the Poisson approximation and an argument similar to the one in Lemma~\ref{lem:ballsnbins}, it is easy to see that when $\sigma \le n$ parties each independently send a flag to a uniformly random filter leaf node out of $n/4$ choices, the probability of the event $E_0$, that there is a leaf node that receives more than $21\log{n}$ flags is less than $n^{-\log\log{n}}$.
	
	Once the flags have been sent to the leaf nodes of the filters, they are forwarded up the filter from nodes to their parents, all the way to the collection node, with the only caveat that nodes do not forward more than $21\log{n}$ flags. Since each node has two children, it follows that each node in the filter receives at most $42\log{n}$ flags, and the same is true of the collection nodes. At the same time, viewing the process as first selecting a collection node, and then a filter leaf node below it, we see as in Lemma~\ref{lem:bigtau} that the probability of the event $E_2$, that there is a filter that receives more than $21s\log{n}$ flags is at most $\frac1{7n^{1.1s}\log{n}}$. Since no node in the filter can get more flags than the filter as a whole, it follows that the filter nodes and the collection nodes all receive no more than $21M\log{n} = \min\{21s\log{n}, 42\log{n}\}$ flags. This shows (b) and the upper bound in (a).
	
	To show that the collection nodes each receives at least $7\log{n}$ flags with high probability, and that together the collection nodes receive at least $\tau$ flags, we will once again track the marked flags. As we have remarked previously, although the marked flags are the first $\tau$ to be sent, by asynchrony, they need not be the first $\tau$ to arrive at the filters.
	Thus, it need not be the case that all these marked flags are forwarded through to the collection nodes. Nevertheless, we will argue that for every marked flag that fails to be forwarded, at least one unmarked flag was forwarded instead.
	To see this, note that as in Lemma~\ref{lem:bigtau}, all the filters receive between $7\log{n}$ and $21\log{n}$ \emph{marked} flags, except with probability $\frac{1+n^{0.9}}{7n^2\log{n}}$. Thus, each node in a filter can have at most $21\log{n}$ marked flags arrive at it.
	
	Now, suppose a filter node fails to forward one or more marked flags. It can only do this if it has previously forwarded $21\log{n}$ flags, and since it can receive at most $21\log{n}$ marked flags, it follows that it has already forwarded at least as many
	\emph{unmarked} flags as it is choosing to ignore marked ones. Once again, by transferring marks from the marked flags that are dropped to the unmarked flags that have been sent in their place, we can ensure that except with probability $\frac{1+n^{0.9}}{7n^2\log{n}}$, between  $7\log{n}$ and $21\log{n}$ marked flags get through each filter to the corresponding collection node, and at least $\tau$ marked flags get through all the filters together, to the collection layer of the first collection subtree. This shows the lower bound in (a) and sets us up to show (c) and (d).
	
	For (c), we will once again pretend, by transferring marks that at each node the marked flags are the first to arrive and be used. As before, we do this without altering the distribution of marked and unmarked flags between collection nodes. Note that each newly marked flag at the collection node corresponds to a distinct
	\oneinput, so the $7\log{n}$ of them used by each of $\tau/(14\log{n})$ collection nodes to generate a \countm message accounts for $\tau/2$
	\oneinputs at the root. This leaves between $0$ and $14\log{n}$ marked flags at each collection node which add up to $\tau/2$ of them across all the collection nodes. Since each collection node forwards up to $14\log{n}$ flags, all the marked flags are forwarded, so that at least $\tau/2$ flags are forwarded to the next collection subtree. Since each of $\tau/(14\log{n})$ collection nodes forwards up to $14\log{n}$ flags, at most $\tau$ flags are forwarded to the next collection tree, proving (d).
	
	Finally, adding up the probabilities of all the bad events we've encountered, we see that for large enough $n$, $\frac{1+n^{0.9}}{7n^2\log{n}}
	+ \frac1{7n^{1.1}\log{n}}+ n^{-\log\log{n}} < \frac1{7n\log{n}}$.
	It follows that with probability at least $1-\frac1{7n\log{n}}$, (a), (b), (c), and (d) are all true, as desired.
	\qed
\end{proof}

We are now ready to study what happens further up in the data structure.
We will say that the algorithm succeeds up to level $j$ if for all $i \le j$ the following are true:
\begin{enumerate}
	\item All the collection nodes in the $i$th collection subtree receive between $7\log{n}$ and $42\log{n}$ flags.
	\item The \countm messages generated in the $i$th subtree account for $\tau/2^i$ \oneinputs at the root.
	\item Between $\tau/2^i$ and $\tau/2^{i-1}$ flags are forwarded from the $i$th collection subtree to the $(i+1)$st collection subtree
\end{enumerate}

\begin{lemma} \label{lem:successj}
	Let $j \le \dcl$.
	In the \alg{Thresh-Count} algorithm, with probability at least $1 - \frac{j}{7n\log{n}}$, the algorithm succeeds up to level $j$.
\end{lemma}
\begin{proof}
	We proceed by induction on $j$.
	We have already established the base case $j=1$ in Lemmas~\ref{lem:bigtau}
	and~\ref{lem:smalltau}.
	Now suppose $j \ge 2$, and for an induction hypothesis we assume that the algorithm succeeds to level $j-1$ with probability at least $1- \frac{j-1}{7n\log{n}}$. Let us condition on this event. This means that between $\tau/2^{j-1}$ and $\tau/2^{j-2}$ flags are forwarded to the $j$th collection subtree, which has $\frac{\tau}{2^{j-1}14\log{n}}$ collection nodes.
	
	Thus, we can apply Lemma~\ref{lem:ballsnbins} with $\alpha$ between $14\log{n}$ and $28\log{n}$. The proof that conditioned on the algorithm having succeeded up to level $j-1$, it succeeds to level $j$, except with probability $\frac{1}{7n\log{n}}$, is identical to the proof of Lemma~\ref{lem:bigtau}. By Bayes' law and the induction hypothesis, the unconditional probability that the algorithm succeeds to level $j$
	\[
	\left(1- \frac{j-1}{7n\log{n}}\right)\left(1-\frac{1}{7n\log{n}}\right) \ge 1-\frac{j}{7n\log{n}},
	\]
	as desired.
	\qed
\end{proof}

\begin{corollary}\label{cor:successroot}
	With probability at least $1 - \frac{1}{7n}$, the root node successfully accounts for at least $\tau$ \oneinputs.
\end{corollary}
\begin{proof}
	The last collection subtree is the one corresponding to $j= \dcl$, and by Lemma~\ref{lem:successj}, with probability at least $1- \frac{\dcl}{7\log{n}}$ the root has accounted for $\sum_{j=1}^{\dcl}\tau/2^j = \tau(1-2^{-\dcl})$
	\oneinputs, and moreover, between $\tau/2^{\dcl}$ and $\tau/2^{\dcl-1}$ flags have been forwarded directly to the root, by the collection nodes in the last collection subtree. Since no randomness is involved, the root eventually receives all of these flags.
	Thus, conditioned on the algorithm succeeding up to level $\dcl$, the root eventually accounts for at least $\tau$ \oneinputs. Since $\dcl < \log \tau <\log n$, the success probability is at least $1-\frac1{7n}$.
	\qed
\end{proof}

We now prove the Theorem~\ref{thm:taucount}.
Lemmas~\ref{lem:bigtau} to~\ref{lem:successj} and Corollary~\ref{cor:successroot} show that with probability at least $1- \frac{1}{7n}$, the root accounts for at least $\tau$ \oneinputs while ensuring the following:
\begin{enumerate}
	\item Filter nodes receive no more than $42\log{n}$ messages and send no more
	than $21\log{n}$ messages.
	\item Collection nodes receive no more than $42\log{n}$ messages and send
	no more than $14\log{n} + 1$ messages. (The extra 1 is for the \countm
	message.)
	\item The root receives no more than $\tau/2^{\dcl-1}= 28\log{n}$ \flag messages.
\end{enumerate}
Additionally, the adding nodes each receive two \countm messages and send one \countm message, and the root receives $\dcl \le \log{n}$ \countm messages, one from each of the collection subtrees.
Individual parties send at most one message each, when their input is set to $\mathbf{1}$. We have already remarked that the messages used in this algorithm can be encoded using two bits.
Thus, in the Up stage of the algorithm each party sends and receives  $O(\log{n})$ messages of constant size.
In the Down stage, \done messages are sent via a canonical complete binary tree, so each party except the root receives exactly one \done message, and each party that is not a leaf in the tree sends (at most) two \done messages.
Since all messages that are sent are eventually received, eventually all the parties receive the \done message and terminate.
Since the depths of the data structure used in the Up stage and the binary tree used in the Down stage are both $\log{n}$, the longest chain of messages is of length $2\log{n}$, and hence the total latency is $O(\log{n})$.
Finally, since the computations done by each node during the algorithm amount to counting the number of messages it receives and generating up to $14 \log{n}$ random numbers, each node performs $O(\log{n})$ computations.
\qed

\subsection{Using Quorums as Nodes in the Count Tree}
So far in this section, we have assumed that all of the nodes in the count tree follow the protocol honestly. However, this is not the case in our MPC model, where some of the parties can play maliciously. To fix this, we assign a quorum to each node in the tree and let the quorums perform the roles of the parties. In our MPC protocol described in Section~\ref{sec:alg}, we introduce Protocol~\ref{pro:waitForInputs} that allows us run the threshold counting algorithm in a malicious setting.
 
Lemma~\ref{lem:qcomm} shows that a quorum $Q$ can securely send a message $M$ to another quorum  $Q^\prime$. However, there is some subtlety involved in using this fact. Every party in a quorum communicates with its parent when it has received at least half as many inputs as the parents' threshold. However, due to asynchrony, multiple messages may arrive simultaneously; when the threshold is set, not all parties in the quorum may be in the same state. Some may already have more inputs than the threshold, while others may still be waiting, because messages from their children have been delayed. Lemma~\ref{lem:qcomm} tells us that if all parties in the quorum send the \emph{same} message to the parent quorum, then the parent quorum can resolve that message.
Thus, in order to ensure that all parties in the quorum send the same message to the parent quorum, we have required that even if a party's received inputs exceed his threshold, it should only inform the parent of having met the threshold, not of having exceeded it. The remaining inputs are held to be sent later.

\section{Asynchronous Quorum Formation} \label{sec:qf}
In this section, we describe the quorum building algorithm of King~\etal~\cite{King:2006:TSS:1170136.1170491,ICDCN11}, and then adapt it to the asynchronous communication model by proving the following theorem:

\begin{theorem}
	Consider $n$ parties connected to each other pairwise in an asynchronous network, where up to \mbox{$t<(\frac14-\epsilon)n$} of them are corrupted, for some small constant $\epsilon>0$. If all honest parties follow the protocol \textsf{Build-Quorums}, then with high probability,
	\begin{enumerate}
		\item the parties agree on $n$ quorums,
		\item each party sends at most $\tilde{O}(\sqrt n)$ field elements,
		\item each party performs $\tilde{O}(\sqrt n)$ computations, and
		\item the protocol latency is $O(\polylog(n))$.
	\end{enumerate}
	\label{thm:quorum-building}
\end{theorem}

One may alternatively use the asynchronous Byzantine agreement protocol of Braud-Santoni~\etal~\cite{Braud-Santoni:2013:FBA:2484239.2484243} to build a set of $n$ quorums. This protocol requires each party on average to send \textsf{polylog}$(n)$ field elements, and perform \textsf{polylog}$(n)$ computations.  However, it is not load-balanced: some parties may send a linear number of field elements. Using this result our MPC protocol needs only logarithmic bits and computations.

We start the description of our protocol by defining the \emph{\rs\ problem}, where the goal is to agree on a single string of length $O(\log{n})$ with a constant fraction for random bits, where for any positive constant $\epsilon$, a $1/2+\epsilon$ fraction of the parties are honest. 
King~\etal~\cite{ICDCN11} present an asynchronous algorithm as an additional result that we call \rsToQ. The \rsToQ\ algorithm can go from a solution for \emph{\rs} problem to the solution for the \emph{quorum building} problem. Thus, their techniques can be extended to the asynchronous model assuming a scalable asynchronous solution for the \rs problem. We describe \textsf{Build-Quorums} algorithm based on \rsToQ\ and an algorithm, that solves \rs problem in the asynchronous model with pairwise channels that we call~\rsAlg.

\begin{algorithm}
	\caption{\textsf{Build-Quorums}}
	\label{pro:cq}
	\medskip
	\algfont
	\textit{Goal.} Generate $n$ quorums.
	
	\begin{enumerate}
		\item All parties run \rsAlg.
		\item All parties run \rsToQ.
	\end{enumerate}
\end{algorithm}

King~\etal~\cite{King:2006:TSS:1170136.1170491} present a synchronous algorithm that a set of parties, up to 1/3 of which are controlled by an adversary, can reach almost-everywhere\footnote{King~\etal~\cite{King:2006:TSS:1170136.1170491} relax the requirement that all honest parties reach agreement at the end of the protocol, instead requiring that a $1- o(1)$ fraction of honest parties reach agreement. They refer to this relaxation as almost-everywhere agreement.} agreement with probability $1 - o(1)$.
Their main technique is to divide the parties into groups of polylogarithmic size; each party is assigned to multiple groups. In parallel, each group uses \binelection algorithm~\cite{Feige} to \emph{elect} a small number of parties from within their group to move on.
This step is recursively repeated on the set of elected parties until size of the remaining parties in this set becomes polylogarithmic. At this point, the remaining parties can solve the \rs problem (similarly, they can run a Byzantine agreement protocol to agree on a bit). Provided the fraction of corrupted parties in the set of remaining parties is less than $1/3$ with high probability, these parties succeed in agreeing on a semi-random string. Then, these parties communicate the result value to the rest of the parties.

Bringing parties to agreement on a semi-random string is trickier in the asynchronous model. The major difficulty is that the \binelection algorithm cannot be used in asynchronous model since the adversary can prevent a fraction of the honest parties from being heard, and then prevent them to be part of the election. We present a similar algorithm to~\cite{King:2006:TSS:1170136.1170491} that solves this issue in asynchronous model with private channels. The main result of this section is as follows.

\begin{theorem}\label{t:mainbyz}
	Suppose there are $n$ parties, for any fix positive $\epsilon$ constant fraction $b < 1/4 - \epsilon $ of which are corrupted. There is a polylogarithmic (in $n$) bounded degree network and a protocol such that:
	\begin{enumerate}
		\item With high probability,  a $1 - O(1/\ln{n})$ fraction of the honest parties agree on the same value (bit or string).
		\item Every honest party sends and processes only a polylogarithmic (in $n$) number of bits.
		\item The number of rounds required is polylogarithmic in $n$.
	\end{enumerate}
\end{theorem}

The important novelty of our method compare to King~\etal~\cite{King:2006:TSS:1170136.1170491} is that instead of \binelection algorithm, we use \hw to decide on the parties who move on to the next level. The simple version of our election method is presented as \simplees in Protocol~\ref{pro:subcommittee} that has the properties described in Lemma~\ref{lem:subcommittee}. The complete protocol and its proof of correctness are given in Section~\ref{sec:cq-proofs}

\begin{algorithm}
	\caption{\simplees}
	\label{pro:subcommittee}
	\medskip
	\algfont
	\textit{Goal.} $\Omega(\ln^8{n})$ parties agree on a subcommittee of size $\Omega(\ln^3{n})$. The protocol is performed by parties $P_1,...,P_k \in W$ with $k = \Omega(\ln^8{n})$.
	
	\begin{enumerate}
		\item Party $P_i$ generate a vector of $c \ln^3{n}$ random numbers chosen uniformly and independently at random from $1$ to $k$ where each random number maps to one party.
		\item Run \hw to compute the component-wise sum modulo $k$ of all the vectors. Arbitrarily, add enough additional numbers from $1$ to $k$ to the sum vector to ensure it has  $c \ln^3{n}$  unique numbers.
		\item Let $W_B$ be the set of winning parties which are those associated with the components of the sum vector.
		\item Return $W_B$ as the elected subcommittee.
	\end{enumerate}
\end{algorithm}

\begin{lemma}\label{lem:subcommittee}
	Let $W$ be a committee of $\Omega(\ln^8{n})$ parties, where the fraction, $f_W$, of honest parties is greater than 3/4. Then, there exists some constant $c$, such that with high probability, the \es protocol elects a subset $W_B$ of $W$ such that $|W_B| = c \ln^3{n}$ and the fraction of honest parties in $W_B$ is greater than $(1-1/ \ln{n})f_W$. The \es protocol uses a polylogarithmic number of bits and polylogarithmic number of rounds in a fully connected network.
\end{lemma}
\begin{proof}
	The proof follows from a straightforward application of union and Chernoff bounds. Let $X$ be the number of honest parties in $W_B$. By the correctness of the \hw algorithm, each party in $W_B$ is randomly chosen from $W$. Let $Y_i$ be an indicator random variable, that equals to 1 if the $i$-th member of $W_B$ is honest. Then, $E[Y_i] = f_W$ and $E[X] = f_Wc_1\ln^3{n}$. Using Chernoff bounds, we have $Pr[X < (1-1/\ln{n})f_Wc_1\ln^3{n}] = Pr[X < (1-1/\ln{n})E[X]] \leq e^{-\frac{E[X]/\ln^2{n}}{2}} < 1/n^c$. Since $f_W>1/2$, setting $c_1 = 4c$, establishes the first part of Lemma \ref{lem:subcommittee}.
	\qed
\end{proof}

We establish a polylogarithmic bound on the number of bits used in \es protocol since the bit cost of \es is polynomial in the number of parties participating in the algorithm.

\subsection{The Election Graph} \label{s:layered network}

Our algorithms make use of an election graph to determine which parties will participate in which elections. This graph was described in~\cite{KSSV,King:2006:TSS:1170136.1170491} and is repeated here.

Before describing the election graph, we first present a result similar to that used in~\cite{Cooper}. Let $X$ be a set of parties. For a collection $\mathcal{F}$ of subsets of $X$, a parameter $\delta$, and a subset $X'$ of $X$, let $\mathcal{F}(X', \delta)$ be the sub-collection of all $F' \in \mathcal{F}$ for which

\[\frac{|F' \bigcap X'|}{|F'|} > \frac{|X'|}{|X|} + \delta.\]

In other words, $\mathcal{F}(X', \delta)$ is the set of all subsets of $\mathcal{F}$ whose overlap with  $X'$ is larger than the ``expected'' size by more than a $\delta$ fraction.
Let $\Gamma(r)$ denote the neighbors of node $r$ in a graph.

\begin{lemma} \label{l:expander}
	Let $l,r,n$ be positive integers such that $l$ and $r$ are all no more than $n$ and $r/l \geq ln^{1-z} n$. Then, there is a bipartite graph $G(L,R)$ such that $|L| = l$ and $|R| = r$ and
	
	\begin{enumerate}
		\item Each node in $R$ has degree $\ln^{z} n$.
		\item Each node in $L$ has degree $O((r/l) \ln^{z}n)$.
		\item Let $\mathcal{F}$ be the collection of sets $\Gamma(r)$ for each $r \in R$. Then, for any subset $L'$ of $L$,\\
		$|\mathcal{F}(L', 1/\ln n )| < \max(l,r)/ \ln^{z-2} n$.
	\end{enumerate}
	
\end{lemma}

The proof of Lemma~\ref{l:expander} follows from a counting argument using the probabilistic method and is omitted.
The following corollaries follows immediately by repeated application of the above lemma.

\begin{corollary} \label{l:expander s-node connection}
	Let $\toplev$ be the smallest integer such that $n/\ln^{\toplev} n \le \ln^{10} n$.
	There is a family of bipartite graphs $G(L_i, R_i), i=0,1,\ldots,
	\toplev$, and constants $c_1$ and $c_2$ such that $|L_{i}| =
	n/\ln^{i} n$, $|R_{i}|=n/\ln^{i+1} n$, and
	
	\begin{enumerate}		
		\item Each node in $R_i$ has degree $\ln^{c_1} n$.
		
		\item Each node in $L_{i}$ has degree $O(\ln^{c_2} n)$.
		
		\item Let $\mathcal{F}$ be the collection of sets $\Gamma(r)$ for each $r \in R$. Then, for any subset $L_i'$ of $L_i$,\\
		$|\mathcal{F}(L_i', 1/\ln n )| < |R_i|/\ln^{6} n$.
		
		\item Let $\mathcal{F'}$ be the collection of sets $\Gamma(l)$ for each $l \in L$. Then, for any subset $R_i'$ of $R_i$,\\
		$|\mathcal{F'}(R_i', 1/\ln n )| < |L_i|/\ln^{6} n$.		
	\end{enumerate}
	
\end{corollary}

\begin{corollary}
	\label{l:expander graph family}  Let $\toplev$ be the smallest integer such that $n/\ln^{\toplev} n \le \ln^{10} n$. There is a family of bipartite graphs $G(L_i, R_i), i=0,1,\ldots, \toplev$, such that $|L_{i}| = n/\ln^{i} n$, $|R_{i}|=n/\ln^{i+1} n$, and
	
	\begin{enumerate}		
		\item Each node in $R_i$ has degree $\ln^5 n$.		
		\item Each node in $L_{i}$ has degree $O(\ln^4 n)$.		
		\item Let $\mathcal{F}$ be the collection of sets $\Gamma(r)$ for each $r \in R$. Then, for any subset $L_i'$ of $L_i$,\\
		$|\mathcal{F}(L_i', 1/\ln n )| < |L_i|/\ln^{3} n$.		
	\end{enumerate}
	
\end{corollary}

Lemma~\ref{l:expander} and its corollaries show there exists a family of bipartite graphs with strong expansion properties which allow the formation of subsets of parties where all but a small fraction contain a majority that are honest.

We are now ready to describe the election graph. Throughout, we refer to nodes of the election graph as \textsf{e-node}s to distinguish them from nodes of the static network. Let $\toplev$ be the minimum integer $\lev$ such that $n/\ln^\lev n \leq \ln^{10} n$; note that $\toplev = O(\ln n/\ln\ln n)$. The topmost layer $\toplev$ has a single \textsf{e-node} which is adjacent to every \textsf{e-node} in layer $\toplev -1$. For the remaining layers $\lev=0,1,...,\toplev-1$, there are $n/\ln^{\lev+1}n$ \textsf{e-node}s. There is an edge between the $i$th \textsf{e-node}, $A$,  in layer $\lev$ and the $j$th \textsf{e-node}, $B$, in layer $\lev+1$ if and only if there is an edge between the $i$th node in $L_{\lev+1}$ and the $j$th node in $R_{\lev+1}$ from Corollary~\ref{l:expander graph family}. In such a case, we say that $B$ is the parent of $A$, and $A$ is the child of $B$. Note that \textsf{e-node}s have many parents.

Each \textsf{e-node} will contain a set of parties known as a {\it committee}. All \textsf{e-node}s, except for the one on the top layer and those in layer 0, will contain $c\ln^3 n$ parties. Initially, we assign the $n$ parties to \textsf{e-node}s on layer $0$ using the bipartite graph $G(L_0,R_0)$ described in Corollary~\ref{l:expander graph family}. The $i^{th}$ party is a member of the committee contained in the $j^{th}$ \textsf{e-node} of layer 0 if and only if there is an edge in $G$ between the $i^{th}$ node of $L_0$ and the $j^{th}$ node of $R_0$. Note every \textsf{e-node} on layer 0 has $\ln^5 n$ parties in it.

The \textsf{e-node}s on higher layers have committees assigned to them during the course of the protocol. Let $A$ be an \textsf{e-node} on layer $\lev>0$, let $B_1,\ldots, B_s$ be the children of $A$ on layer $\lev-1$, and suppose that we have already assigned committees to \textsf{e-node}s on layers lower than $\lev$. If $\lev < \toplev$, we assign a committee to $A$ by running \es on the parties assigned to $B_1,\ldots,B_s$, and assigning the winning subcommittee to $A$. (Note that we can run each of these elections in parallel.)  If $A$ is at layer $\toplev$, the parties in $A$, $B_1,\ldots,B_s$, run \heavyba\ for Byzantine agreement.

\subsection{Static Network with Polylog-Bounded Degree} \label{s: static network}

We now repeat the description of the bounded degree static network~\cite{King:2006:TSS:1170136.1170491} and show how it can be used to hold elections specified by the election graph. For each \textsf{e-node} $A$, we form a collection of parties which we call it \textsf{s-node}: $s(A)$. Intuitively, the \textsf{s-node} $s(A)$ serves as a central communication point for an election occurring at \textsf{e-node} $A$. 
Our goal is to bound the fraction of \textsf{s-node}s controlled by the adversary by a decreasing function in $n$, namely $ 1/\ln^{10} n$, for each layer. As the number of \textsf{s-node}s grows much smaller with each layer, we need to make each \textsf{s-node} more robust. To do this, the number of parties contained in the \textsf{s-node} increases with the layer. Specifically, the \textsf{s-node}s for layer $i$ are sets of $\ln^{i+12} n $ parties. We determine these \textsf{s-node}s using the bipartite graph from Lemma \ref{l:expander}, where $L$ is a collection of $n$ nodes, one for each party, $R$ is the set of \textsf{s-node}s for layer $i$ and the degree of each node in $R$ is set to $\ln^{i+12} n$. The neighbors of each node in $R$ constitute a set of parties in an \textsf{s-node} on layer $i$.

We use the term {\em link} to denote a direct connection in the static network. The communications for an election $A$ will all be routed through $s(A)$: a message from a party $x$ to $s(A)$ on layer $i$ will pass from the party to a layer $0$ \textsf{s-node}, whose parties will forward the message to a layer $1$ \textsf{s-node} and so on, the goal being to reliably transmit the message via increasingly larger \textsf{s-node}s up to $s(A)$.
Similarly, communications to an individual party $x$ from $s(A)$ will be transmitted down to a layer $0$ \textsf{s-node} whose parties will transmit the message to $x$. We describe the connections in the static network.

\begin{description}
	\item[Connections in the static network.] Consider the following:
	\begin{itemize}	
		\item Let $A$ be an \textsf{e-node} on layer $0$ in the election graph.
		Every party in $A$ has a link to every party in $s(A)$.
		
		\item Let $A$ and $B$ be \textsf{e-node}s in the election graph at levels $i$ and $i-1$ respectively such that $A$ is a parent of $B$. Thus, $s(A)$ has $\ln^{i+12} n$ parties in it and $s(B)$ has $\ln^{i+11} n$ parties in it. Let $G$ be a bipartite graph as in Lemma~\ref{l:expander} where $L$ is the set of parties in $s(A)$, $R$ is the set of parties in $s(B)$ and the degree of $R$ is set to $\ln^{c_1} n$ and the degree of $L$ is set to $O(\ln^{c_2}n)$.  If there is an edge between two nodes in $L$ and $R$ respectively, then the corresponding party in $s(A)$ has a link to the corresponding party in $s(B)$. We will sometimes say that $s(A)$ is adjacent to $s(B)$ in the static network.
	\end{itemize}
\end{description}

The following lemma follows easily from the application of Lemma \ref{l:expander} and its corollaries.
Item (1) follows from Lemma 3.1; items (2) and (4)  from Corollary 3.2; and item (3) from Corollary 3.1. Although item (2) only makes a guarantee about layer 0 \textsf{e-node}s, we will see eventually that with high probability, the fraction of corrupted \textsf{e-node}s on every layer is small.

\begin{lemma}
	With high probability, the election graph and the static network have the following properties:
	
	\label{l:network properties}
	\begin{enumerate}
		\item (Bad \textsf{s-node}s) Any \textsf{s-node} whose fraction of corrupt parties exceeds $b+ 1/\ln n$ will be called {\it bad}. Else, we will call the \textsf{s-node} {\it good}. No more than a $1/\ln^{10} n$ fraction of \textsf{s-node}s on any given layer are bad.
		
		\item (Bad \textsf{e-node}s) Any \textsf{e-node} whose fraction of corrupt parties exceeds $b+ 1/\ln n$ will be called {\it bad}. Else we call the \textsf{e-node} {\it good}. No more than a $1/\ln^{2} n$ fraction of
		\textsf{e-node}s on layer 0 are bad.
		
		\item  (Bad \textsf{s-node} to \textsf{s-node} connection)
		For any pair of \textsf{e-node}s $A$ and $B$ joined in the election graph, the parties in \textsf{s-node}s $s(A)$ and $s(B)$ are linked such that the following holds. For any subset $W_A$ of parties in $s(A)$, at most a $1/\ln^6 n$ fraction of parties in $s(B)$ have more than a $|W_A|/|s(A)| + 1/\ln n$ fraction of their links to $s(A)$ with parties in $W_A$.
		
		\item  (Bad \textsf{e-node} to \textsf{e-node} connection) Let $|I|$ represent the total number of \textsf{e-node}s on layer $i$ in the election graph.
		For any set $W$ of \textsf{e-node}s on any layer $i$,  at most a $1/\ln^2 n$ fraction of \textsf{e-node}s on layer $i+1$ have more than $|W|/|I| + 1/\ln n$ fraction of their neighbors in $W$.
		
	\end{enumerate}
\end{lemma}

The degree of the static network is polylogarithmic.

\subsection{Communication Protocols} \label{s:low level communication}

A {\em permissible path} is a path of the form $P=x, s(A_0), s(A_1),...,.s(A_i)$ where $x$ is a party in $A_0$, $i$ is the current layer of elections being held, each $A_j$ is an \textsf{e-node} on layer $j$, and there is an edge in the election graph between $A_j$ and $A_{j+1}$ for $j=0,...,i$. Each party $y$ in an \textsf{s-node} $s(A)$ on each layer $j$ keeps a {\em List} of permissible paths that determine which parties' messages it will forward. The \lst\ (for $y\in s(A)$) represents $y$'s view of which parties are elected (to the subcommittee) at $A$ that are still participating in elections on higher layers. If $y$'s \lst\ indicates that $x$ is such a party, then the \lst\ will also have the entire path for $x$, which stretches from $x$ to the elections on layer $i$ in which $x$ is currently participating in. We have the following definitions.

\begin{itemize}
	\item We say a {\em \textsf{s-node} knows a message} [resp., {\em knows a permissible path}, or resp., {\em knows a \lst\ of permissible paths}] if $1-b-2/\ln n$ parties in the \textsf{s-node} are honest and receive the same message [resp., have the same path on their \lst s, or resp., all have the same \lst.]
	
	\item A permissible path $P$ is good if every \textsf{s-node} on the path knows $P$. Else the path is bad. We will show our construction of the static network ensures at most a $1/\ln n$ fraction of the permissible paths are bad.
\end{itemize}

We now describe three primitive communication subroutines: \Sendhop, \Send, and \MessagePass. The subroutine \Sendhop\ describes how \textsf{s-node}s (with direct links) communicate with each other, \Send\ describes how a party communicates with an \textsf{s-node}, and \MessagePass\ describes how two parties communicate with each other.

\begin{description}
	\item[\Sendhop$(s, r, m,P)$.]
	A message $m$ can be passed from $s$ (the sender) to $r$ (the receiver) from a level $i$ to a level $i-1$ or from a level $i$ to a level $i+1$, where $s$ and $r$ are \textsf{s-node}s on these layers or one of $s, r$ is a 0-layer \textsf{s-node} and the other is a party. If a party $x$ sends a message to a layer $0$ \textsf{s-node} $s(A)$ it sends the message to every party in $s(A)$ (note by construction it will have a direct link with every party in $s(A)$). Similarly if a message is sent from a layer $0$ \textsf{s-node} $s(A)$ to a party $x$, every party in $s(A)$ sends the message to $x$.
\end{description}

When an \textsf{s-node} $s(A)$ sends a message to \textsf{s-node} $s(B)$, every party in $s(A)$ sends the message to those parties of $s(B)$ to which it has a direct link. When each party in $s(B)$ receives such a set of messages, it waits until it receives the same messages from the majority to determine the message. If there is no majority value, the party ignores the message. Along with sending the message the parties also send information which specifies along which path $P$ the message is being sent. Each time a message is received by a party of an \textsf{s-node} $s(B)$ on layer $j
\leq i$, it checks that:

\begin{enumerate}
	\item The message came from the \textsf{s-node} previous to it in the path $P$; if not the message is dropped.
	\item The path $P$ (or its reverse) is on its \lst\ of permissible paths. If not, the message is dropped.	
	\item Only messages that conform to the protocol in size and number are forwarded up and down the permissible paths. If more or longer messages are received from a party, messages from that party are dropped.
\end{enumerate}

\begin{description}
	\item[\Send$(s,r,m,P)$.]
	Of the first two parameters, one must be a party (``$x$'') and one must be an \textsf{s-node} (``$s(A)$''). The path $P$ contains the first parameter $s$ as its start and the second parameter $r$ as its endpoint. \Send$(s,r,m,P)$ sends the message $m$ from $s$ to $r$ along the path $P$ via repeated application of \Sendhop.
	
	\item[\MessagePass$(x\in A, y\in B, m, P_x, P_y)$.]
	Both $A$ and $B$ are adjacent \textsf{e-node}s. Hence, $s(A)$ and $s(B)$ are adjacent in the static network.
	A message from party $x$ in \textsf{e-node} $A$ sends message $m$ to party $y$ in
	\textsf{e-node} $B$ by first calling \Send$(x, s(A),m,P_x)$. Then, $s(A)$ sends $m$ to $s(B)$ by calling \Sendhop$(s(A),s(B),m,P)$, where $P$ is the path consisting of two \textsf{s-node}s $s(A),s(B)$. Finally, the message is transmitted from $s(B)$ to $y$ by calling \Send$(s(B), y, m, P_y^r)$, where $P_y^r$ is the reversal of path $P_y$.
\end{description}

\subsection{\rsAlg Protocol} \label{sec:srs-agreement}

Before describing the \rsAlg protocol, we first adapt the \es protocol for the static network. Let $A$ be an \textsf{e-node} with children $B_1,\ldots,B_s$, and let $X$ be the set of all parties from $B_1,\ldots,B_s$. For each $i\in[s]$ and $x \in B_i$, let $P_x$ denote a good path of \textsf{s-node}s from $x$ to $s(B_i)$ concatenated with $s(A)$. At the start of the election for $A$, we assume that each node in $P_x$ knows $P_x$ and $s(A)$ knows $\{P_x~|~ x \in X\}$. 

We now describe the implementation of the \es algorithm. Every party $x\in X$  generate a vector of random numbers chosen uniformly and independently at random where each random number maps to one party. The parties use the \hw protocol to determine the winners 
(recall that the number of parties in \textsf{e-node}s is always polylogarithmic, so this can be done sending only polylogarithmic messages). The list of winners is sent up to $s(A)$, where each party in $s(A)$ takes a majority to determine the winners. Then, $s(A)$ sends down the list of winners along all the permissible paths to each party $x\in X$.
Parties on the path (\ie, in \textsf{s-node}s along the path) update their \lst s of permissible paths to remove those party-paths who lost as well as those party-pairs who won too many elections (we will quantify this shortly), and make $\ln^4 n$ copies of each of the winners' paths and concatenate a different layer $i +1$ \textsf{s-node} parent onto each one. We present a detailed description of \es in the following.
\begin{algorithm}
	\caption{\es}
	\medskip
	\algfont
	\textit{Goal.} Adapted version of \simplees for static networks.
	\medskip
	\begin{enumerate}
		\item For each  $x \in X$: {\em   // This stage done in parallel}
		\item
		\begin{indentpar}{0.5cm}
			Party $x$ randomly selects one of $k/(c_1\ln^3 n)$ random numbers chosen uniformly and independently at random from zero to $k$ where each random number maps to one party.
		\end{indentpar}
		\item Parties in $X$ run \hw to compute the component-wise sum modulo $k$ of all the vector. Arbitrarily, add enough additional numbers to the vector to ensure it has $c\ln^3{n}$ unique numbers.
		\item Let $M$ be the set of winning parties, which are those associated with some component of the vector sum.
		\item
		\begin{indentpar}{0.5cm}
			Each $y \in X$ sends $M$ to $s(A)$ by calling \Send$(y, s(A),M, P_y)$.
		\end{indentpar}
		\item
		\begin{indentpar}{0.5cm}
			Parties in $s(A)$ determine $M$ by waiting until they receive the majority of same messages. These become the elected parties.
		\end{indentpar}
		\item For each party $x \in X$ that is elected, the parties in $s(A)$ use  \Send$(s(A), x, m, P^r_x)$ to tell $x$, along with each \textsf{s-node} in $P_x$, that $x$ was elected.
		\item Each party in each \textsf{s-node} revises its list of permissible paths to:
		\begin{indentpar}{0.5cm}
			Retain only the winners.
			Eliminate parties who have won more than 8 elections.
			Make $\ln^4{n}$ copies of remaining permissive paths, concatenating each with a different \textsf{s-node} neighbor on layer $i + 1$.
		\end{indentpar}
		\item $s(A)$ sends its list to every adjacent \textsf{s-node} $s(B)$ on layer $i + 1$ using \Sendhop$(s(A), s(B), m, P)$, where $P$ is the path consisting only of $s(A)$, $s(B)$.
	\end{enumerate}
\end{algorithm}

The condition in Step $5$ that requires parties who have won more than $8$ elections to be eliminated is a technical condition that insures the protocol is load-balanced and parties in an \textsf{s-node} do not communicate more than a polylogarithmic number of bits. We now describe the \rsAlg protocol.
\begin{algorithm}
	\caption{{\sf Scalable-}\rsAlg}
		\medskip
		\algfont
		\textit{Goal.} Parties agree on a semi-random string.
	\begin{enumerate}
		\item For $l = 1$ to $l^*$:
		\item
		\begin{indentpar}{0.3cm}
			For each \textsf{e-node} $A$ in layer $l$, let $B_1, . . . ,B_s$ be the children of $A$ in layer $l-1$ of the election graph, and
		\end{indentpar}
		\item
		\begin{indentpar}{0.6cm}
			If $l < l^*$, run \es on the parties in nodes $B_1, . . . ,B_s$. Assign winning parties to node $A$.
		\end{indentpar}
		\item
		\begin{indentpar}{0.6cm}
			Else parties in nodes $B_1, . . . ,B_s$ solve \rs problem.
		\end{indentpar}
		\item Let $A^*$ be the \textsf{e-node} on layer $l^*$, every party $x$ assigned to $A^*$ communicates the result of Step 4 to $s(A^*)$ using \Send$(x, s(A^*), m, P_x)$.
		\item Every party in $s(A^*)$ waits for the majority of same message to determine the result of Step 4.
	\end{enumerate}
\end{algorithm}

Since every party is a member of $s(A^{*})$, steps $5$ and $6$ will insure the final result of the protocol is communicated to every party.

\subsection{Proof of \textsf{Build-Quorums}} \label{sec:cq-proofs}
To establish the correctness of the protocol presented in Section~\ref{sec:srs-agreement}, we first state some claims regarding the primitive communication protocols. Their proofs follow by straightforward probabilistic arguments and are omitted in the interest of space. Recall the fraction of corrupted parties is $b$, where $b < 1/4 - \epsilon$ for any fix positive $\epsilon$.

\begin{claim} \label{c:nodeToNode}
	Let $s(A)$ and $s(B)$ be \textsf{s-node}s on consecutive layers. Assume the following conditions hold:
	\begin{enumerate}		
		\item Both $s(A)$ and $s(B)$ are good.
		
		\item $s(A)$ is on a permissible path known by $s(B)$.
		
		\item There exists a set $W$ of parties from $s(A)$ such that for every message $m$,  all parties in $W$ are honest and agree on a message $m$. Further $W$ consists of at least a $1-b-2/\ln n$ fraction of the parties in $s(A)$.
	\end{enumerate}
	
	Then, there is a set $W'$ of parties from $s(B)$ such that for every message $m$, every party in $W'$ is honest and agrees on the message $m$ after \Sendhop$(s(A),s(B),m,P)$ is called. (Here, $P$ is the path $s(A),s(B)$.)  Further, $W'$ consists of all but a $1/\ln^6 n$ fraction of the honest parties in $s(B)$.
\end{claim}
\begin{proof}
	Every party in $W$ is honest and sends the same massage to its connected parties in $s(B)$. The parties in $s(B)$ can afford to wait for the majority of same messages, since $s(A)$ is good and $W$ consists of at least $1-b-2/\ln{n}$ fraction of parties which is more than $1/2$ and for majority we need to receive a fraction of $1/2$ same messages from the parties in $s(A)$. Thus, all honest party but a $1/\ln^6{n}$ fraction of parties in $s(B)$ will eventually receive the message based on corollary \ref{l:expander s-node connection}.
	\qed
\end{proof}

\begin{claim}\label{c:partyToNode}
	Let $x$ be an honest party. Assume $P_x$ is a good path.
	Then, after \Send$(x, s(A), m, P_x)$ is executed, there is a fixed set $W$ of honest parties which contains all but a $1/\ln^6 n$ fraction of the honest parties in $s(A)$ and every party $z \in W$ agrees on $m$.	
\end{claim}

An election at \textsf{e-node} $A$ is {\em legitimate} if the following two conditions hold simultaneously for more than a $3/4$ fraction of parties $x$ participating in the election at $A$: (1) party $x$ is honest; (2) The path $P_x$ is good.

\begin{lemma}
	\label{l:legitimateElections}
	For a legitimate election at node $A$, let $X$ be a set of honest parties with good permissible paths. (Note $|X| >
	3\ln^8 n/4$.) Let $W$ be the set of honest parties in $s(A)$ that know $X$. Then, after the execution of {\sc Elect-subcommittee}, the parties in $W$ know the winners of the election in $A$, as do the \textsf{s-node}s that belong to good paths $P_x$.
\end{lemma}
\begin{proof}
	From Claim~\ref{c:partyToNode}, we have that every message $m$ sent by \MessagePass$(y \in B_i, z \ \in B_j, m,P_y, P_z)$ from $y\in X$ to $z \in X$ is received by some fixed set $W$ of honest parties in $s(B_i)$, such that $W$ contains at least $1-b-2/\ln n$ fraction of the parties in $s(B_i)$.
	By Claim~\ref{c:nodeToNode}, every message sent by $y$ is received by $z$. Since $X$ contains more than $3/4$ of the total parties participating in the election, (after running \hw) all the parties in $X$ will all agree on the same set of for random parties. Thus, after the parties in $X$ send these values to $s(A)$, $s(A)$ will know the winners. When $s(A)$ sends these winners to $X$, by repeated application of Claim~\ref{c:nodeToNode}, we have every $x \in X$ and every \textsf{s-node} in $P_x$ will know these winners.
	\qed
\end{proof}

We have shown that in a legitimate election at node $A$, $s(A)$ knows the list of winners. We next consider when paths are dropped from the permissible path \lst s.

\subsubsection{Permissible Paths Removal}
\ Let $y$ be a party in some \textsf{s-node} on layer $i$. A permissible path $P_x$ is removed from a party $y$'s \lst\ on layer $i$ if $y$ receives a message from an \textsf{s-node} above it in $P_x$, indicating either $x$ has won more than 8 elections or $x$ lost in the election held at the last node of $P_x$. Here, we consider when $P_x$ is removed for the former reason, \ie, we give an upper bound on the fraction of parties that are reported to have won too many elections on layer $i$.

First we consider the effect of legitimate elections. The following lemma, a version of which appears in \cite{KSSV,King:2006:TSS:1170136.1170491}, shows that on a given layer a very small fraction of honest parties win more than 8 times in legitimate elections.

\begin{lemma}
	With high probability, the parties that win more than 8 elections, counting multiplicities, account for no more than a $16/\ln^3 n$ fraction of the honest parties that are winners of legitimate elections.
\end{lemma}

Next, we bound the effect of elections that are not legitimate. We first consider the case where $s(A)$ is good, yet the fraction of honest parties participating in $A$ with good paths is less than 3/4. For the remainder of the proof we shall treat such an \textsf{e-node}\  $A$ as a bad \textsf{e-node}.

\begin{claim}
	\label{c:bad election}
	Suppose less than a $1/7$ fraction of the honest parties of a good $s(A)$ agree on a message $m$. Then, after
	\Sendhop$(p(A),p(B), m, P)$ is executed, all but a $1/\ln^{6} n $ fraction of the honest parties in $s(B)$ will ignore $m$.
\end{claim}

\noindent\begin{proof}
	Even if the corrupted parties agree on $m$, since $b <1/4$, the total fraction of parties in $s(A)$ sending the message $m$ is less than 11/28. Thus, at most a $1/\ln^{6} n$ fraction of the parties in $s(B)$ will receive $m$ from a majority of parties in $s(A)$.
	\qed
\end{proof}

Hence, a good $s(A)$ can only communicate with seven different sets of winners to the \textsf{s-node}s below it. Since each honest party will send $\ln^3 n$ winners, the total number of winners sent is at most $7 \ln^3 n$. Therefore, a bad \textsf{e-node} can cause at most $7
\ln^3 n$ parties to have their permissible paths removed.

Next, we consider the effect of a bad \textsf{s-node}. We will assume one bad \textsf{s-node} $s(A)$ on layer $i$ can cause the removal of all the permissible paths for every party participating in the election at $A$. Since $\ln^8 n$ parties participate in an election, and fewer than a $1/\ln^{10} n$ fraction of the \textsf{s-node}s are bad on a layer, the fraction of honest winners affected is less than $1/\ln^2 n$. Thus, we can bound the fraction of the honest winners on any layer $i$ that have their permissible paths removed by $1/\ln ^2 n+ 1/\ln^3 n + 7\beta_{i}$; where $\beta_{i}$ represents the fraction of bad \textsf{e-node}s on layer $i$.
Thus, we have the following lemma.

\begin{lemma}
	\label{l:win too often} Assume the fraction of bad \textsf{e-node}s on layer $i$ is bounded by $c/\ln^2 n$, for some constant $c$. Then, the fraction of honest winners that have their permissible paths removed on layer $i$ is bounded by $8c/\ln^2 n$.
\end{lemma}

\subsubsection{Proof of Theorems~\ref{t:mainbyz}}

We now complete the proof of Theorem~\ref{t:mainbyz} which follows from the following lemma.

\begin{lemma}
	\label{l:main}
	On layer $i$, with high probability, at least a $1-4/\ln^2 n$ fraction of \textsf{s-node}s $s(A_j)$ have the following properties:
	\begin{enumerate}
		\item $s(A_j)$ is good.

		\item At least a $1-b- 4i /\ln n$ fraction of the parties in node $A_j$ are honest and have good paths to $s(A_j)$ (note this implies $s(A_j)$ knows this path). That is, $A_j$ is a good \textsf{e-node}.
	\end{enumerate}
\end{lemma}
\begin{proof}
	We prove the lemma by induction. On all layers and particularly layer 0, only a $1/\ln^{10} n$ fraction of the \textsf{s-node}s are bad. If $s(A)$ is good, then every party in $A$ has a good path to $s(A)$. Further by construction all but a $1/\ln^2 n$ fraction of the \textsf{e-node}s on layer 0 consist of at least a $1-b - 1/\ln n$ fraction of honest parties. So the lemma is true on layer 0.
	
	Assume the lemma is true for layer $i$. Then, a $1- 4/\ln^2 n$ fraction of \textsf{e-node}s are good, more specifically these \textsf{e-node}s have at least a $1 - b - 4i/\ln n$ fraction of honest parties that have a good path to their corresponding \textsf{s-node}. Since the election is legitimate by Lemmas~\ref{lem:subcommittee}
	and~\ref{l:legitimateElections}, with high probability, after \es at least a $1 - b - 4i/\ln n - 1/\ln n$ fraction of the parties elected are honest and have a good path to any good parent of their \textsf{s-node}. Thus, at least a $1 - b - (4i + 1)/\ln n$ fraction of the parties elected at layer $i$ are honest and have good paths to good parent \textsf{s-node}s on layer $i+1$. By Lemma~\ref{l:win too often} this fraction is reduced by at most $32/\ln^2 n$. Thus, at least a $1 - b - (4i + 2)/\ln n$ fraction of the parties elected at layer $i$ are honest and have good paths to good parent \textsf{s-node}s on layer $i+1$. Since the fraction of bad \textsf{s-node}s on layer $i+1$ is at most $1/\ln^{10} n$, by Corollary ~\ref{l:expander graph family} at least a $1-1/\ln^2n - 1/\ln^{10} n$ fraction of the
	\textsf{e-node}s (and their corresponding \textsf{s-node}s) are good on layer $i+1$, and have at least a $1 - b - (4i + 2)/\ln n - 1/\ln n$ fraction of honest parties that have good paths to their corresponding \textsf{s-node}s.
	\qed
\end{proof}

By Lemma~\ref{l:main}, with high probability the layer $\toplev$ \textsf{e-node} is good.
Thus, the parties in this \textsf{e-node} succeed in solving the \rs problem (Step 4 of algorithm \rsAlg). Since all the parties are in the \textsf{s-node} (though they may appear multiple times) corresponding to $A$ on $\toplev$, by Claim~\ref{c:partyToNode} all but a $O(1/\ln n)$ fraction of the honest parties learn the final result.
To prove the number of bits sent by each party is polylogarithmic we note each party is in a polylogarithmic number of \textsf{e-node}s and \textsf{s-node}s on each layer $i$, and participates in at most a polylogarithmic number of election on layer $i$. Since the number of layers is $O(\ln n)$ Theorem~\ref{t:mainbyz} follows. Finally, the correctness of Theorem~\ref{thm:quorum-building} follows from Theorem~\ref{t:mainbyz} and the correctness of \rsToQ protocol.

\section{Conclusion} \label{sec:end}
We described a Monte Carlo algorithm to perform asynchronous MPC in an scalable manner. Our protocols are scalable in the sense that they require each party to send $\tilde{O}(m/n + \sqrt n)$ messages and perform $\tilde{O}(m/n + \sqrt n)$ computations. They tolerate a static adversary that controls up to a $\fbad-\epsilon$ fraction of the parties, for $\epsilon$ any positive constant. We showed that our protocol is secure in the universal composability framework.
We also described efficient algorithms for two important building blocks of our protocol: threshold counting and quorum building. These algorithms can be used separately in other distributed protocols.

The following problems remain open. Can we prove lower bounds for the communication and computation costs for Monte Carlo MPC?  Can we implement and adapt our algorithm to make it practical for a MPC problem such as the beet auction problem described in~\cite{bogetoft2009secure}. Finally, can we prove upper and lower bounds for resource costs to solve MPC in the case where the adversary is \emph{adaptive}, able to take over parties at any point during the algorithm?

\section*{Acknowledgements}
The authors would like to acknowledge supports from NSF under grants CCF-1320994, CCR-0313160, and CAREER Award 644058. We are also grateful for valuable comments from Ran~Canetti (Boston University), Shafi~Goldwasser (MIT), Aniket~Kate (Saarland), Yehuda~Lindell (Bar-Ilan), and Seth~Pettie (UMich).

\bibliographystyle{alpha}
\bibliography{security}

\end{document}